\documentclass[12pt]{article}
\usepackage{amsfonts}
\usepackage{amssymb}
\usepackage{amsbsy}

\input epsf.sty

\newcommand{\BEQ}{\begin{equation}}     % Gleichungen Anfang ..
\newcommand{\BEA}{\begin{eqnarray}}
\newcommand{\EEQ}{\end{equation}}       % .. und Ende
\newcommand{\EEA}{\end{eqnarray}}
\newcommand{\eps}{\varepsilon}          % epsilon
\newcommand{\vph}{\varphi}              % rundes phi
\newcommand{\g}{{\mathfrak{g}}}         % einige deutsche Buchstaben fuer 
\newcommand{\h}{{\mathfrak{h}}}         % die Liealgebren
\newcommand{\s}{{\mathfrak{s}}}
\newcommand{\n}{{\mathfrak{n}}}
\newcommand{\m}{{\mathfrak{m}}}

\newcommand{\al}{\alpha}

\newcommand{\half}{{1\over 2}}
\newcommand{\Del}{\Delta}

\newcommand{\R}{\mathbb{R}}
\newcommand{\C}{\mathbb{C}}
\newcommand{\D}{{\rm d}}                % gerades d fuer Ableitungen
\newcommand{\II}{{\rm i}}               % gerades i fuer komplexe Einheit
\newcommand{\tr}{{\rm tr\ }}            % Spursymbol
\newcommand{\dddot}[1]{\stackrel{...}{#1}} % dreifache Ableitung 
\newcommand{\wit}[1]{\widetilde{#1}}    % weite Schlange
\renewcommand{\vec}[1]{\boldsymbol{#1}} % Vektoren fettgedruckt
\newcommand{\appsection}[2]{\setcounter{equation}{0} \section*{Appendix #1. #2}
\renewcommand{\theequation}{#1\arabic{equation}}
              \renewcommand{\thesection}{#1} }
\catcode`\@=11
\def\numberbysection{\@addtoreset{equation}{section}
        \def\theequation{\thesection.\arabic{equation}}}
\numberbysection

\begin{document}

\begin{titlepage}

%%{\hfill \tt \today; pr\'eliminaire !}

\vskip 1.5 cm
\begin{center}
{\Large \bf Schr\"odinger invariance and space-time symmetries}
\end{center}

\vskip 2.0 cm
   
\centerline{ {\bf Malte 
Henkel}$^a$\footnote{courriel: {\tt henkel@lpm.u-nancy.fr}} and 
{\bf J\'er\'emie 
Unterberger}$^b$\footnote{courriel: 
{\tt Jeremie.Unterberger@antares.iecn.u-nancy.fr}}}
\vskip 0.5 cm
\centerline {$^a$Laboratoire de Physique des 
Mat\'eriaux,\footnote{Laboratoire associ\'e au CNRS UMR 7556} 
Universit\'e Henri Poincar\'e Nancy I,} 
\centerline{ B.P. 239, 
F -- 54506 Vand{\oe}uvre l\`es Nancy Cedex, France}
\vskip 0.5 cm
\centerline {$^b$Institut Elie Cartan,\footnote{Laboratoire 
associ\'e au CNRS UMR 7502} D\'epartement de Math\'ematiques, 
Universit\'e Henri Poincar\'e Nancy I,} 
\centerline{ B.P. 239, 
F -- 54506 Vand{\oe}uvre l\`es Nancy Cedex, France}

~\vskip 1.5 cm 

\begin{abstract}
\noindent
The free Schr\"odinger equation with mass $\cal M$ can be turned into a
non-massive Klein-Gordon equation via Fourier transformation with respect to
$\cal M$. The kinematic symmetry algebra $\mathfrak{sch}_d$ of the free
$d$-dimensional Schr\"odinger equation with $\cal M$ fixed appears
therefore naturally as a parabolic subalgebra of the complexified
conformal algebra $\mathfrak{conf}_{d+2}$ in $d+2$ dimensions. The explicit
classification of the parabolic subalgebras of $\mathfrak{conf}_3$ yields
physically interesting dynamic symmetry algebras. This allows us to
propose a new dynamic symmetry group relevant for the description
of ageing far from thermal equilibrium, with a dynamical exponent $z=2$. 
The Ward identities resulting from the invariance under 
$\mathfrak{conf}_{d+2}$ and its parabolic subalgebras are derived and the 
corresponding free-field energy-momentum tensor is constructed. We also
derive the scaling form and the causality conditions for the two- and 
three-point functions and their relationship
with response functions in the context of Martin-Siggia-Rose theory.
\end{abstract}

\end{titlepage}

%%%%%%%%%%%%%%%%%%%%%%%%%%%%%%%%%%%%%%%%%%%%%%%%%%%%%%%%%%%%%%%%%%%%%%%%%%%%%%%%
\section{Introduction}
%%%%%%%%%%%%%%%%%%%%%%%%%%%%%%%%%%%%%%%%%%%%%%%%%%%%%%%%%%%%%%%%%%%%%%%%%%%%%%%%

The Schr\"odinger group {\sl Sch}$(d)$ is defined as the following set 
of space-time transformations in $d$ space dimensions
\BEQ \label{1:gl:SCH1}
t \longmapsto t' = \frac{\alpha t + \beta}{\gamma t + \delta} \;\; , \;\;
\vec{r} \longmapsto \vec{r}' = \frac{{\cal R} \vec{r} + \vec{v} t + \vec{a}}
{\gamma t + \delta} \;\; ; \;\; \alpha\delta - \beta\gamma =1
\EEQ
 where $\cal R$ is a rotation matrix. Consider the free Schr\"odinger equation
\BEQ \label{1:gl:Schr}
2\m \II\frac{\partial \phi}{\partial t}-
\frac{\partial^2 \phi}{\partial \vec{r}^2} =
2{\cal M}\frac{\partial \phi}{\partial t} -
\frac{\partial^2 \phi}{\partial \vec{r}^2} =0
\EEQ
where the mass $\m$ is a constant. In 1972, Niederer \cite{Nied72} 
showed that  the maximal kinematic invariance group of  (\ref{1:gl:Schr})
is the Schr\"odinger group {\sl Sch}$(d)$, while Lie had already shown in 1882
that the free diffusion equation is invariant under 
{\sl Sch}$(d)$ \cite{Lie1882}. 
The action of {\sl Sch}$(d)$ on the space of solutions
$\phi$ of (\ref{1:gl:Schr}) is projective, that is,
the wave function $\phi=\phi(t,\vec{r})$ 
transforms into
\BEQ
\phi(t,\vec{r}) \longmapsto \left( T_g \phi\right)(t,\vec{r}) =
f_g[g^{-1}(t,\vec{r})] \phi[g^{-1}(t,\vec{r})]
\EEQ
(the companion function $f_g$ is explicitly known \cite{Nied72}). 
Independently, Hagen \cite{Hage72} showed that the free-field action from which
(\ref{1:gl:Schr}) can be derived as equation of motion is 
Schr\"odinger-invariant. In this paper, we shall mainly consider the
Lie algebra $\mathfrak{sch}_d$
of the Schr\"odinger group {\sl Sch}$(d)$. In particular, one has the
following set of generators for  $\mathfrak{sch}_1$:
\BEA 
X_{-1} = -\partial_t \;\; , \;\; Y_{-1/2} = -\partial_r & &
\mbox{\rm time and space translations} \nonumber \\
Y_{1/2} = - t\partial_r - {\cal M} r & &
\mbox{\rm Galilei transformation} \nonumber \\
X_{0} = -t\partial_t - \frac{1}{2} r\partial_r - \frac{x}{2} & &
\mbox{\rm dilatation} \nonumber \\
X_{1} = -t^2\partial_t - tr\partial_r - \frac{\cal M}{2} r^2 - 2 x t & &
\mbox{\rm special Schr\"odinger transformation} \nonumber \\
M_{0} = - {\cal M} & &
\mbox{\rm phase shift}
\EEA
Here, ${\cal M}=\II \m$ and $x$ is the scaling 
dimension of the wave function $\phi$ on which
the generators of $\mathfrak{sch}_1$ act.
 For a solution of the free Schr\"odinger
equation (\ref{1:gl:Schr}) one has $x=d/2$.  
The non-vanishing commutators of $\mathfrak{sch}_1$ are 
\BEQ
\left[ X_n, X_{n'} \right] = (n-n') X_{n+n'} \;\; , \;\;
\left[ X_n, Y_m \right] = \left(\frac{n}{2}-m\right) Y_{n+m} \;\; , \;\;
\left[ Y_{1/2}, Y_{-1/2}\right] = M_{0} 
\EEQ  
($n,n'\in\{\pm 1,0\}$, $m\in\{\pm 1/2\}$). 
The  Schr\"odinger group has been introduced \cite{Nied72,Hage72} 
as a non-relativistic 
analogue of the conformal group in $d$ dimensions 
(whose Lie algebra will be denoted here $\mathfrak{conf}_d$). 
Indeed, it was
argued by Barut \cite{Baru73} that the Schr\"odinger Lie algebra
$\mathfrak{sch}_d$ could be obtained from the conformal algebra
$\mathfrak{conf}_{d+1}$ {\it ``\ldots by a combined process of group
contraction and a `transfer' of the transformation of mass to 
the co-ordinates''}. The
projective unitary irreducible representations of  
the Schr\"odinger group {\sl Sch}($d$) 
are classified in \cite{Perr77}. In particular,  
{\sl Sch}($1$) is isomorphic to  the semi-direct product 
{\sl Sl}($2,\R$)$\ltimes$ {\sl H}($1$), where {\sl Sl}($2,\R$) 
comes from the exponentiation of the $X$-generators, 
and the $1$-dimensional Heisenberg group {\sl H}($1$) 
from the exponentiation of $Y_{\pm 1/2}$ and $M_0$.  

Schr\"odinger invariance has been considered in a wide variety of 
situations, for example celestial mechanics \cite{Duva91}, non-relativistic
field theory \cite{Berg92,Burd72,Duva86,Haug94,Jack90a,Jack90b,Mehe00} and/or 
non-relativistic quantum mechanics \cite{Ghos01,Herr02,Jack72}, hydrodynamics
\cite{Hass00,Ivas97,Jahn01,ORai01} or dynamical scaling 
\cite{Henk92,Henk94,Henk94a}, see also references therein. 
Our interest in this dynamical symmetry comes from the consideration of
situations of dynamical scaling such that the correlators of field operators
$\phi_i$ transform covariantly under dilatations (with $b$  constant)
\BEQ \label{1:gl:cov}
\langle \phi_1 (b^z t_1,b \vec{r}_1) \cdots 
\phi_n (b^z t_n ,b \vec{r}_n )\rangle =b^{-x_1-\ldots -x_n }
\langle \phi_1 (t_1 ,\vec{r}_1 )\cdots  
\phi_n (t_n ,\vec{r}_n)\rangle
\EEQ 
where $z$ is the dynamical exponent. Such a dynamic scaling behaviour occurs in 
many physical situations, for example critical dynamics or else in the 
phase-ordering kinetics undergone by a spin system quenched from a disordered 
initial state to a temperature $T<T_c$, where $T_c>0$ is the critical 
temperature (see e.g. \cite{Bray94,Cate00} for reviews). Eq.~(\ref{1:gl:cov}) 
is  compatible with
Galilei invariance only if $z=2$. By analogy with conformal invariance 
\cite{Bela84}, one may ask whether a generalization of (\ref{1:gl:cov}) to a
{\em local scale invariance} with space-time-dependent rescaling
factors $b=b(t,\vec{r})$ is possible. Indeed, it has been shown recently
that infinitesimal generators of local scale transformations with any
given value of $z$ can be constructed \cite{Henk02}. In turn, admitting
local scale invariance as a hypothesis of dynamics leads to explicit
expressions for two-point functions which can be tested in specific models.
These phenomenological predictions have so far been confirmed at the Lifshitz 
points of spin systems with competing interactions \cite{Henk97,Plei01} 
and in the non-equilibrium ageing behaviour of ferromagnetic spin systems 
\cite{Cala02,Cann01,Godr00a,Godr00b,Godr02,Henk01,Henk02a,Henk02b,Pico02}.

These `experimental' confirmations provide some credibility to the
idea of local scale invariance. However, an understanding of the origin
of local scale invariance is still lacking. For example, in the
present tests of local scale invariance the values of the scaling 
dimensions $x_i$ are still considered as free parameters. An 
approach similar to the one of $2D$ conformal invariance as initiated in
\cite{Bela84} and which allows, among other things, 
to obtain the $x_i$ from the representation theory of the Virasoro algebra, 
to find all $n$-point functions and furthermore to list the universality
classes, is not available (see e.g. \cite{diFr97,Henk99}
for introductions to conformal invariance). As a first step in this
direction, we shall undertake here a case study of the simplest case, namely
Schr\"odinger invariance, which is realized for $z=2$ \cite{Henk94}.  

In conformal field theory, the central object is the energy-momentum tensor
and the main tool the conformal Ward identities it satisfies. In order to
prepare an analogous study for a Schr\"odinger-invariant system, we shall
go here through an exercise in classical non-relativistic field theory. 
In Galilei-invariant field theories, a technical problem comes from the
fact that wave functions transform under {\em projective} 
representations (as given by the companion function $f_g$ and parametrized 
by the non-relativistic mass $\m$ \cite{Nied72}) rather than under true
representations. However, Giulini \cite{Giul96} pointed out that by treating
the mass $\m$ as a dynamical variable and going over to the new field $\psi$
defined by (we merely write here the single-particle case)
\BEQ
\phi(t,\vec{r}) = \frac{1}{\sqrt{2\pi}} 
\int_{\mathbb{R}} \!\D\zeta\, e^{-\II\m\zeta} \psi(\zeta,t,\vec{r})
\EEQ
one obtains a true unitary representation $\overline{T}_g\psi$ of the 
Galilei group (see \cite{Giul99} for a discussion on the interpretation
of the Bargmann superselection rule). 
In section 2, we shall show that his result generalizes to the
full Schr\"odinger group {\sl Sch}($d$) as well as to a certain 
infinite-dimensional extension thereof. 

Treating the mass $\m$ as a dynamical variable from the outset allows
further insights. In section 3, we shall derive a precise relationship 
between the Schr\"odinger algebra and the Lie algebra of the conformal group. 
In particular, we shall show
that for the complexified Lie algebras, one has 
$\mathfrak{sch}_d\subset\mathfrak{conf}_{d+2}$. We shall show that the 
maximal kinematic invariance group of the $d$-dimensional free Schr\"odinger 
equation with {\em variable} mass is the $(d+2)$-dimensional conformal group.
We also discuss the relevance of the parabolic subalgebras of 
$(\mathfrak{conf}_{d+2})_{\C}$ for physical applications, notably to ageing
in simple magnets. Finally, we correct the claims of Barut and show that his
would-be group contraction from $\mathfrak{conf}_{d+1}$ to $\mathfrak{sch}_d$
should rather be viewed as a projection from $(\mathfrak{conf}_{d+2})_{\C}$ to 
a new subalgebra $\wit{\mathfrak{alt}}_{d}$ distinct from $\mathfrak{sch}_d$.
In section 4, we consider the Schr\"odinger Ward identities which must
be satisfied by the energy-momentum tensor. The improved energy-momentum
tensor satisfying the resulting symmetry requirements on the level of
classical field theory will be explicitly constructed, for theories
built either from fields $\phi(t,\vec{r})$ 
or from fields $\psi(\zeta,t,\vec{r})$. We also show how to generalize these
considerations such as to make them applicable to ageing phenomena, where
time-translation invariance no longer holds. In section 5, we discuss the 
resulting two- and three-point functions and show that they satisfy the
causality conditions required for their physical interpretation as
response functions. We conclude in section 6. In appendix~A,
we provide details on the non-relativistic limit of the conformal algebra. 
In appendix~B, we derive the causality conditions satisfied by the two- and
three-point functions. Appendix~C reviews basic facts about parabolic 
subalgebras.

%%%%%%%%%%%%%%%%%%%%%%%%%%%%%%%%%%%%%%%%%%%%%%%%%%%%%%%%%%%%%%%%%%%%%%%%%%%%%%%%
\section{Extended Schr\"odinger transformations}
%%%%%%%%%%%%%%%%%%%%%%%%%%%%%%%%%%%%%%%%%%%%%%%%%%%%%%%%%%%%%%%%%%%%%%%%%%%%%%%%

We consider the following infinite-dimensional
extension of $\mathfrak{sch}_1$, denoted by ${\cal S}_1^{\infty}$, 
and spanned by the generators $\{X_n, Y_m, M_n\}$ 
with $n\in\mathbb{Z}$ and $m\in\mathbb{Z}+\frac{1}{2}$. They
are of the following form \cite{Henk94}
\BEA
X_n &=& -t^{n+1}\partial_t - \frac{n+1}{2}\ t^n r\partial_r 
- (n+1)\frac{x}{2} t^n
-\frac{n(n+1)}{4}{\cal M} t^{n-1} r^2 \nonumber \\
Y_m &=& -t^{m+1/2} \partial_r - \left( m+\frac{1}{2}\right) t^{m-1/2} r{\cal M}
\nonumber \\
M_n &=& - {\cal M} t^n
\label{gl:Schr}
\EEA
Here, the constant $x$ is the scaling dimension of the field $\phi(t,r)$ 
on which these generators act, see below. 
The generators satisfy the following non-vanishing commutation relations
\BEA
\left[ X_n, X_{n'} \right] &=& (n-n') X_{n+n'} \;\; , \;\;
\left[ X_n, Y_m \right] = \left(\frac{n}{2}-m\right) Y_{n+m} \;\; , \;\;
\nonumber \\
\left[ X_n, M_{n'} \right] &=& -n' M_{n+n'} \;\; , \;\;
\left[ Y_m, Y_{m'}\right] = (m-m') M_{m+m'}. 
\label{2:SKomm}
\EEA
Extensions to $d>1$ are straightforward, see  \cite{Henk02}. The special
case ${\cal M}=0$ of ${\cal S}_1^{\infty}$ was rediscovered later as a kinematic
symmetry of the $1D$ Burgers equation \cite{Ivas97}. 
In particular, the $X_n$ satisfy the commutation relations of the Virasoro
algebra without central charge. Indeed, one might consider the generators 
$X_n$, $Y_m$ and $M_n$ as the components of associated conserved currents. 
Then the conformal dimensions of these currents, as measured by $X_0$, are 
$\dim X=2$, $\dim Y=\frac{3}{2}$ and $\dim M=1$. In other terms, just as the
finite-dimensional Schr\"odinger algebra was a semi-direct product of
$\mathfrak{sl}(2,\R)$ with a Heisenberg algebra, the  
Lie algebra ${\cal S}_1^{\infty}$
is a semi-direct product of a Virasoro algebra without central charge 
(extending $\mathfrak{sl}(2,\R)$) and of
a {\it two-step nilpotent} (that is to say, whose brackets are central)
Lie algebra generated by the $Y_m$ and $M_n$, 
extending the Heisenberg algebra.

We shall assume that the action of the generators (\ref{gl:Schr})  
describes the
transformation of a non-relativistic field $\phi(t,r)$ of mass $\cal M$. It
is straightforward to integrate the infinitesimal transformations through
formal exponentiation. From the generators $X_n$ we find the following 
coordinate transformations (where $\beta(t')$ is a non-decreasing function 
of $t'$)  
\BEQ
t = \beta(t') \;\; , \;\; r = r' \sqrt{ \dot{\beta}(t')}
\EEQ
Here and in the sequel the dot will denote the derivative with respect to 
the time variable. 
The field $\phi$ transforms into $\phi'$ such that
\BEQ
\phi(t,r) = \dot{\beta}(t')^{-x/2}\, \exp\left( -\frac{\cal M}{4} 
\frac{\ddot{\beta}(t')}{\dot{\beta}(t')}{r'}^2\, \right) \phi'(t',r')
\EEQ
The infinitesimal generator $X_n$ in eq.~(\ref{gl:Schr}) is recovered from 
$\beta(t)=t-\eps {t}^{n+1}$ in the limit $\eps\to 0$.
 
Similarly, exponentiation of $Y_m$ gives the coordinate transformations
\BEQ
t = t' \;\; , \;\; r = r' -\alpha(t)
\EEQ
and the field transforms as 
\BEQ
\phi(t,r) = \exp\left( {\cal M}\left(\frac{1}{2}\alpha(t')\,
\dot{\alpha}(t') -r'\,\dot{\alpha}(t')\right)\right) \phi'(t',r')
\EEQ
The infinitesimal generator $Y_m$ is recovered from 
$\alpha(t)=-\eps {t}^{m+1/2}$ in the limit $\eps\to 0$.
Finally, exponentiation of $M_n$ merely changes the phase of $\phi$ 
\BEQ
\phi(t,r) = \exp \left({\cal M}\gamma(t)\right) \phi'(t,r) 
\EEQ
without any changes in the coordinates. 

In the sequel, it will be useful to work with the Fourier transform of the
field and of the generators with respect to $\cal M$. We define the new
field $\psi$ as follows 
\BEQ \label{2:gl:phipsi}
\phi(t,r) = \frac{1}{\sqrt{2\pi}} 
\int_{\mathbb{R}} \!\D\zeta\, e^{-\II {\cal M}\zeta}\, \psi(\zeta,t,r)
\EEQ
As a consequence of (\ref{1:gl:Schr}), this field satisfies the equation of
motion, provided $\lim_{\zeta\to\pm\infty}\psi(\zeta,t,r)=0$
\BEQ \label{2:gl:Schr}
2\II \frac{\partial^2 \psi}{\partial t\partial\zeta} + 
\frac{\partial^2 \psi}{\partial r^2} =0
\EEQ
which for the sake of brevity we shall also call a Schr\"odinger equation.

The generators of ${\cal S}_1^{\infty}$ act on the 
Fourier transform $\psi(\zeta,t,r)$ as follows:    
\BEA
X_n &=& -t^{n+1}\partial_t - \frac{n+1}{2}\ t^n r\partial_r 
- (n+1)\frac{x}{2} t^n
+\II \frac{n(n+1)}{4} t^{n-1} r^2 \partial_{\zeta} \nonumber \\
Y_m &=& -t^{m+1/2} \partial_r + \II \left( m+\frac{1}{2}\right) t^{m-1/2}
 r \partial_{\zeta}
\nonumber \\
M_n &=&  \II t^n \partial_{\zeta}
\label{gl:Schr2}
\EEA
Let us write out for later use the action on $\psi$ of those generators of 
$\mathfrak{sch}_1$ which contain a phase shift 
\BEA 
Y_{1/2} = - t\partial_r + \II r \partial_{\zeta} & &
\mbox{\rm Galilei transformation} \nonumber \\
X_{1} = -t^2\partial_t - tr\partial_r + {\II\over 2} r^2 
\partial_{\zeta} -  x t & &
\mbox{\rm special Schr\"odinger transformation} \nonumber \\
M_{0} = \II \partial_{\zeta} & &
\mbox{\rm phase shift}
\EEA

It turns out that the field $\psi$ transforms in a simpler way than the
field $\phi$ considered above. The complicated phases acquired by the
field $\phi$ are replaced by a translation of the new internal coordinate
$\zeta$, and at most a scaling factor remains. 
Under the action of the $X_n$, we find
\BEQ \label{3:gl:Xnbeta}
t = \beta(t') \;\; , \;\; r = r' \sqrt{ \dot{\beta}(t')} \;\; , \;\;
\zeta=\zeta'-{\II\over 4} {\ddot{\beta}(t')\over \dot{\beta}(t')} r'^2
\EEQ
and the field $\psi$ transforms into $\psi'$ such that
\BEQ
\psi(\zeta,t,r) = \dot{\beta}(t')^{-x/2}\,  \psi'(\zeta',t',r')
\EEQ
Similarly, exponentiation of $Y_m$ gives the coordinate transformations
\BEQ \label{3:gl:Ymalpha}
t = t' \;\; , \;\; r = r' -\alpha(t)\;\; , \;\; \zeta=\zeta'
+\II r'\, \dot{\alpha}(t')-{\II\over 2}\alpha(t')\, \dot{\alpha}(t')
\EEQ
and the field $\psi(\zeta,t,r)=\psi'(\zeta',t',r')$ transforms trivially. 
The absence of a phase factor under the Galilei transformation $Y_{1/2}$
(where $\alpha(t)=vt$) was observed before by Giulini \cite{Giul96}. 

Finally, the $M_n$ merely give time-dependent translations in the 
coordinate $\zeta$.

Summarizing, we have shown: the 
algebra ${\cal S}_{1}^{\infty}$ acts on fields $\phi(t,r)$ with a fixed
mass as a projective representation. Conjugation
by Fourier transformation with respect to $\cal M$ changes this into a true 
representation of the same algebra, now acting on functions $\psi(\zeta,t,r)$.

%%%%%%%%%%%%%%%%%%%%%%%%%%%%%%%%%%%%%%%%%%%%%%%%%%%%%%%%%%%%%%%%%%%%%%%%%%%%%%%%
\section{Relation between Schr\"odinger and conformal transformations}
%%%%%%%%%%%%%%%%%%%%%%%%%%%%%%%%%%%%%%%%%%%%%%%%%%%%%%%%%%%%%%%%%%%%%%%%%%%%%%%%

In this section, we shall investigate the relationship between the 
Schr\"odinger Lie algebra $\mathfrak{sch}_1$ and the conformal 
Lie algebra $\mathfrak{conf}_{3}$. 

We start from the three-dimensional massless Klein-Gordon equation
\BEQ \label{3:gl:KG}
\partial_{\mu} \partial^{\mu} \Psi(\vec{\xi}) = 0 
\EEQ
The Lie algebra of the maximal kinematic group of this equation is the 
conformal algebra $\mathfrak{conf}_{3}$, with generators
\BEA
P_{\mu} &=& \partial_{\mu} \nonumber \\
M_{\mu\nu} &=& \xi_{\mu}\partial_{\nu} - \xi_{\nu}\partial_{\mu}\nonumber\\
K_{\mu} &=& 2 \xi_{\mu}\xi^{\nu}\partial_{\nu} - 
\xi_{\nu}\xi^{\nu}\partial_{\mu} +2x \xi_{\mu} \label{3:gl:konfG} \\
D &=& \xi^{\nu}\partial_{\nu} +x \nonumber 
\EEA
($\mu,\nu\in\{-1,0,1\}$) 
which represent, respectively, translations, rotations, special transformations
and the dilatation. Here $x$ is the scaling dimension of the field $\psi$.
If $\psi$ is a solution of (\ref{3:gl:KG}), $x=1/2$. 

It is well-known that ${\mathfrak {conf}}_3$ is 
isomorphic to ${\mathfrak{so}}(4,1)$. An
explicit identification is given by the ${\mathfrak{so}}(4,1)$ generators 
${\cal J}_{\mu \nu}$ as follows, where now $-1\leq \mu,\nu\leq 3$
\BEQ
{\cal J}_{\mu \nu}=\II M_{\mu \nu}\;\; , \;\;
{\cal J}_{2,3}=\II D \;\; , \;\; 
{\cal J}_{2 \mu}={-{\II\over 2}} (P_{\mu}+K_{\mu})\;\; , \;\; 
{\cal J}_{3 \mu}={-{\II\over 2}} (P_{\mu}-K_{\mu}).  
\EEQ

Next, we define the physical coordinates as\footnote{We use the short-hand
$\sqrt{\II\,}=e^{\II\pi/4}$ throughout.}  
\BEQ
t=\half (-\xi_0+\II \xi_{-1}) \;\; , \;\;
\zeta=\half (\xi_0+\II \xi_{-1}) \;\; , \;\;
r=\sqrt{\frac{\II}{2}\,} \xi_1
\EEQ 
and $\psi(\zeta,t,r)=\Psi(\vec{\xi})$. Then the Klein-Gordon
equation (\ref{3:gl:KG}) reduces to the Schr\"odinger equation (\ref{2:gl:Schr})
\BEQ
\left( 2\II \frac{\partial^2}{\partial \zeta\partial t} 
+ \frac{\partial^2}{\partial r^2} \right) \psi(\zeta,t,r) 
= 0.
\EEQ
It follows that the generators of ${\mathfrak{sch}}_1$ are linear 
combinations (with complex coefficients) of the above ${\mathfrak{conf}}_3$ 
generators, so that we have an inclusion of  the complexified Lie algebra   
$({\mathfrak{sch}}_1)_{\C}$ into $({\mathfrak{conf}}_3)_{\C}$. Explicitly 
\BEA
X_{-1}=\II (P_{-1}-\II P_0) \;\; , \;\;
X_0=-\half D+{\II\over 2} M_{-1 0}\;\; , \;\; 
X_1=-{\II\over 4} (K_{-1}+\II K_0) 
\nonumber\\
Y_{-1/2}=-\sqrt{\frac{2}{\II}\,} P_1 \;\; , \;\; 
Y_{\half}=-\sqrt{\frac{\II}{2}\,} (M_{-1 1}+\II M_{0 1})\;\; , \;\; 
M_0=P_{-1}+\II P_0.
\EEA
Four additional generators should be added in order to 
get the full conformal Lie algebra $({\mathfrak{conf}}_3)_{\C}$. 
We take them in the following form 
\BEA
N =\II M_{-1 0}=-t\partial_t +\zeta \partial_{\zeta} & &  
\mbox{\rm time-phase symmetry} 
\nonumber \\
V_- = -\sqrt{\frac{\II}{2}\,} (M_{-1 1}-\II M_{0 1})= 
-\zeta \partial_r+\II r\partial_t & &  
\mbox{\rm ``dual'' Galilei transformation} 
\nonumber \\
W = -{\II\over 4} (K_{-1}-\II K_0)=
-\zeta^2 \partial_{\zeta}-\zeta r\partial_r+{\II\over 2} r^2 \partial_t
-x\zeta & & 
\mbox{\rm ``dual'' special transformation} \\
V_+ = -\sqrt{\frac{\II}{2}\,}K_1= 
-2tr\partial_t-2\zeta r\partial_{\zeta}-(r^2+2\II\zeta t)\partial_r-2xr 
\nonumber & &  
\mbox{\rm transversal inversion}.
\EEA
The generators $V_-$ and $W$ are, up to constant coefficients, the 
complex conjugates of $Y_{1/2}$ and $X_1$, respectively, in the 
coordinates $\xi^{\mu}$, hence their names. The complex conjugation becomes
the exchange $t\leftrightarrow\zeta$ in the physical coordinates $(\zeta,t,r)$.

\begin{figure}
\epsfxsize=80mm
\centerline{\epsffile{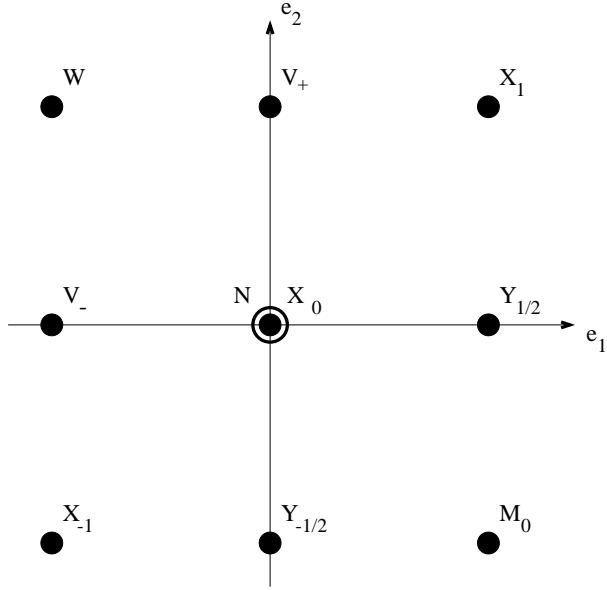}}
\caption{ Roots of $B_2$ and their relation with the generators of the
Schr\"odinger algebra $\mathfrak{sch}_1$. The double circle in the centre
denotes the Cartan subalgebra $\mathfrak{h}$. 
\label{Abb1}}
\end{figure}

In order to understand these results, we recall a few basic facts from the
theory of Lie algebras \cite{Knap86}. 
The Lie algebra $({\mathfrak{conf}}_3)_{\C}$ is simple
and of rank 2. The  generators $D$ and $N$ span a  Cartan sub-algebra.
 We now show that the generators of 
${\mathfrak{sch}}_1$ are {\em root vectors} -- in other words, they are 
common eigenvectors of the commuting generators $N$ and $D$. Let
$e_1$ and $e_2$ be the linear forms on $\R N\oplus\R D$ defined 
by $e_i(N)=\delta_{i,1}$ and $e_i(D)=\delta_{i,2}$.  
Recall that $\lambda^{\mu} e_{\mu}$ is a {\it root} if there is
a non-zero element $Z_{\lambda}$ of $({\mathfrak{conf}}_3)_{\C}$ 
such that $[N,Z_{\lambda}]= \lambda_1 Z_{\lambda}$ and     
$[D,Z_{\lambda}]=\lambda_2 Z_{\lambda}$. Since our complexified
conformal algebra is isomorphic to ${\mathfrak{so}}(5,\C)$, its set of roots
$\Delta$ is of type $B_2$. One finds that
$\Delta=\Delta_+ \cup \Delta_-$, where
\BEQ \label{3:gl:posWur}
\Delta_+=\{- e_2,e_1+ e_2,e_1,e_1-e_2\} \;\; , \;\;  
\Delta_-=-\Del_+=\{ e_2,-(e_1+ e_2),-e_1,-e_1+e_2\},
\EEQ
The elements of $\Delta_+$ are called positive roots and the elements of
$\Delta_-$ are called negative roots.  
The root vectors can be identified explicitly 
\BEA
X_{-e_2}=Y_{-1/2},\  X_{e_1+e_2}=X_1,\  X_{e_1}=Y_{1/2},\  X_{e_1-e_2}=M_0, 
\nonumber \\
X_{e_2}=V_+, \ X_{-(e_1+e_2)}=X_{-1}, \ X_{-e_1}=V_-,\  X_{-(e_1-e_2)}=W.
\label{3:gl:rootX}
\EEA
These results are summarized in figure~\ref{Abb1}. Each of the points in
the diagram indicates a root space. They are labelled by the corresponding
generators from $({\mathfrak{conf}}_3)_{\C}$, according to 
eq.~(\ref{3:gl:rootX}).

%%==============================================================================
%%==============================================================================

We may use this information to list some interesting
subalgebras of $({\mathfrak{conf}}_3)_{\C}$, using the notion of
{\it parabolic} subalgebras explained in appendix~C.
If two root spaces $Z_1, Z_2$ have coordinates $(i_1,j_1), (i_2,j_2)$
in figure~\ref{Abb1}, then the root space $[Z_1,Z_2]$ 
will have coordinates $(i_1+i_2,j_1+j_2)$. 
Therefore, subalgebras may be easily obtained by taking a subset of
the points in figure~\ref{Abb1} which is invariant under
the addition of coordinates. Furthermore, the {\it Weyl group} of the
conformal Lie algebra $({\mathfrak{conf}}_3)_{\C}$ is given by the
discrete set of transformations $(e_1,e_2)\mapsto (\pm e_1,\pm e_2)$ or
$(e_1,e_2)\mapsto (\pm e_2,\pm e_1)$. On the physical coordinates,
they can be implemented by the action of an element of the conformal group
 and will therefore 
give isomorphic (conjugate) subalgebras.

The Weyl group is generated by $w_1: (e_1,e_2)\mapsto (-e_2,-e_1)$ and
$w_2: (e_1,e_2)\mapsto (-e_1,e_2)$. Both appear as simple symmetries on
figure 1 below.  On the physical coordinate space, these
act as $(\zeta,t,r)\mapsto(\zeta',t',r')$ such that 
\BEA
w_1 : & & \zeta=\zeta'+{\II r'^2\over 2t'} \;\; , \;\;
t=-1/t'\;\; , \;\; r=r'/t'
\nonumber \\
w_2 : & & \zeta= t' \;\; , \;\; t = \zeta' \;\; , \;\; r = r'
\EEA
The inversion $w_1$ belongs to the Schr\"odinger group as can be checked from
eq.~(\ref{3:gl:Xnbeta}), while the duality $w_2$ is a conformal transformation.

We arrive this way at the following list of standard  parabolic subalgebras,
see appendix~C for details. 
There are three of them, up to conjugation. We
shall also mention at the same time natural subalgebras of these, with one
or two generators of the Cartan subalgebra removed, that contain essentially
the same information: 
\begin{enumerate}
\item
the algebra $\mathfrak{age}_1$ generated by $X_0,Y_{-1/2},M_0,Y_{1/2},X_1$. 
Adding the generator $N$, we obtain the  algebra 
$\wit{\mathfrak{age}}_1 := \mathfrak{age}_1 \oplus \mathbb{C} N$, which is the
{\it minimal standard parabolic subalgebra} (see appendix~C),
namely, it is the sum
of the  Cartan subalgebra and of the positive root spaces.
It is also possible to dismiss 
altogether the generator $X_0$ or replace it by any linear combination 
of $X_0$ and $N$.
\item
the Schr\"odinger algebra $\mathfrak{sch}_1$. One may also add to it the 
generator $N$. We call 
$\wit{\mathfrak{sch}}_1 :={\mathfrak{sch}}_1\oplus \mathbb{C} N$ the 
parabolic subalgebra thus obtained.
\item
the algebra $\mathfrak{alt}_1$ generated by $D,Y_{-1/2}, M_0,Y_{1/2},V_+,X_1$.
As before, one may add the generator $N$ and obtains thus the
parabolic subalgebra
$\wit{\mathfrak{alt}}_1 := \mathfrak{alt}_1 \oplus \mathbb{C} N$.
\end{enumerate}
Note that the algebras $\widetilde{\mathfrak{sch}}_1$ and 
$\wit{\mathfrak{alt}}_1$ are maximal non-trivial sub-algebras
of $({\mathfrak{conf}}_3)_{\C}$, as expected from the theory of parabolic
subalgebras. This is easily checked 
from the commutators. On the other hand, $\wit{\mathfrak{age}}_1$ is the
intersection of $\widetilde{\mathfrak{sch}}_1$ and 
$\wit{\mathfrak{alt}}_1$. In the first case, the generator $X_{-1}$ is taken
out and in the second case, the generator $V_+$.

It may be interesting on physical grounds to introduce also 
the images of these algebras
under the Weyl symmetry $(e_1,e_2)\mapsto (e_1,-e_2)$. 
This gives the following  new subalgebras:
\begin{enumerate}
\item[4.]
the algebra generated by $M_0,Y_{1/2},V_+,X_1$ and any linear 
combination of $X_0$ and $N$ (or both);
\item[5.]
the algebra generated by $X_0+N,M_0,Y_{1/2},W,V_+,X_1$; one may also 
add the generator $N$.
\end{enumerate}
 
The consideration of these subalgebras might yield physically interesting 
applications. For example, it has been recently shown that the response
functions of simple ferromagnetic systems undergoing ageing after a
quench from some initial state to a temperature below the equilibrium
critical temperature $T_c$ are determined from their covariance under the
infinitesimal transformations contained in $\mathfrak{age}_d$ 
\cite{Henk01,Henk02}. Since ageing phenomena do 
break time-translation invariance generated by $X_{-1}$, a subalgebra without
this generator is needed. The existence of the subalgebra $\mathfrak{alt}$
suggests that the response functions might also transform covariantly
under the conformal generator $V_+$. 

A new type of application  is suggested by the fourth and fifth subalgebras. 
In contrast to $\mathfrak{sch}_1$, not only time-translation 
invariance ($X_{-1}$), 
but also space translation invariance ($Y_{-1/2}$) is broken. 
The breaking of both of these
would be a necessary requirement to describe ageing processes in disordered,
e.g. glassy systems. It remains to be seen whether the larger algebra, with
both $V_+$ and $W$ present, or  the smaller one with only $V_+$, is 
realized in physical systems. Tests of this possibility are currently being 
performed and will be described elsewhere. 
 
Another interesting way (motivated by an analogy with the scheme of
group contractions) to look at how these subalgebras 
sit inside the conformal algebra is to consider a family of linear maps
depending on a parameter $c$, that gives in the $c\to\infty$ limit 
a kind of projection of ${\mathfrak{conf}}_3$
onto any of these subalgebras. This is easy to 
construct by using conjugation by $\exp\big( (aN+bD)\log c\big)$ for adequate 
$a,b$, which multiplies each of the root vectors above by a certain power of 
$c$. For instance, in the new coordinates 
\BEQ
\zeta'=c^2 \zeta, t'=t, r'=cr,
\EEQ 
any root vector with coordinates $(i,j)$ in figure~\ref{Abb1} is multiplied
by $c^{i-j}$, so the $X$-generators and $N$ are preserved, $Y_{\pm 1/2}$ 
are multiplied by $c$, $M_0$
by $c^2$, while the other generators go to zero. So, in a certain way, 
one has a projection $\mathfrak{conf}_3\to\wit{\mathfrak{sch}}_1$.
Similarly, if one sets 
\BEQ
\zeta'=c^{2+\mu}\zeta, t'=c^{-\mu}t, r'=cr\ \ (\mu>0),
\EEQ 
then $X_{\pm 1}$ is multiplied by $c^{\pm\mu}$, $X_0$ and $N$ remain constant,
and $Y_{- 1/2}$, $Y_{1/2}$ and $M_0$ are multiplied, respectively,
by $c, c^{(1+\mu)}$ and $c^{2+\mu}$, while the other generators go to zero. 
Therefore, in the $c\to\infty$ limit, one has a projection 
$\mathfrak{conf}_3\to\wit{\mathfrak{age}}_1$. In physical applications,
one usually considers families of $c$-dependent maps such that 
$c$ can be interpreted as the speed of light. 
Indeed, Barut \cite{Baru73} had claimed that a group contraction from 
$\mathfrak{conf}_{d+1}$ to $\mathfrak{sch}_d$ were possible. We shall 
reconsider his argument in appendix~A. Working with the masses as dynamical
variables from the very beginning, we shall show that his procedure rather
gives a map $\mathfrak{conf}_3\to\wit{\mathfrak{alt}}_1$ in the 
non-relativistic limit.

One may also form the infinite-dimensional algebra 
$\wit{\cal S}_1^{\infty}:={\cal S}_1^{\infty}\oplus\mathbb{C} N$, 
which has besides eq.~(\ref{2:SKomm}) the following non-vanishing commutators
\BEQ
\left[ X_n, N \right] = n X_n \;\; , \;\;
\left[ Y_m, N \right] = \left(m+\frac{1}{2}\right) Y_m \;\; , \;\;
\left[ M_n, N \right] = (n+1) M_n
\EEQ
Note that, both in $\widetilde{\mathfrak{sch}}_1$ and in 
$\wit{\cal S}_1^{\infty}$, the
generator $M_0$ is no longer central.

In order to see which of these generators form indeed a symmetry
algebra of the free Schr\"odinger equation (\ref{2:gl:Schr}), consider the
$1D$ Schr\"odinger operator
\BEQ
{\cal S} = 2\II \partial_{\zeta}\partial_t + \partial_r^2
\EEQ
It is straightforward to check that 
\BEA
\left[ {\cal S}, X_n \right] &=& 
-(n+1) t^n {\cal S} -\II n(n+1) \left( x-\frac{1}{2}\right) 
t^{n-1} \partial_{\zeta} -\frac{n^3-n}{2} t^{n-2} r^2\, \partial_{\zeta}^2 
\nonumber \\
\left[ {\cal S}, Y_m \right] &=& 
-2\left( m^2 -\frac{1}{4}\right) t^{m-3/2} r\, 
\partial_{\zeta}^2 
\nonumber \\
\left[ {\cal S}, M_n \right] &=& -2n t^{n-1} \partial_{\zeta}^2 
\nonumber \\
\left[ {\cal S}, V_{-} \right] &=& 0 \:=\: \left[ {\cal S}, N \right]
\nonumber \\
\left[ {\cal S}, V_{+} \right] &=& 2(1-2x)\partial_t -4r{\cal S} 
\nonumber \\ 
\left[ {\cal S}, W \right] &=& \II(1 - 2x)\partial_r - 2\zeta {\cal S}
\label{3:gl:Kommu}
\EEA
Therefore, under the action of the conformal algebra $(\mathfrak{conf}_3)_{\C}$
any solution $\psi$ of the Schr\"odinger equation ${\cal S}\psi=0$ 
with a scaling dimensions $x=1/2$ is mapped onto another 
solution of (\ref{2:gl:Schr}). 
Restricting to the subalgebra $\mathfrak{sch}_1$, we recover the well-known
invariance of the free Schr\"odinger equation with 
fixed mass \cite{Nied72,Hage72}. 

The results of this section can be extended to an arbitrary space 
dimension $d$, but we shall not discuss this here.

%%%%%%%%%%%%%%%%%%%%%%%%%%%%%%%%%%%%%%%%%%%%%%%%%%%%%%%%%%%%%%%%%%%%%%%%%%%%%%%%
\section{Schr\"odinger-invariant free fields}
%%%%%%%%%%%%%%%%%%%%%%%%%%%%%%%%%%%%%%%%%%%%%%%%%%%%%%%%%%%%%%%%%%%%%%%%%%%%%%%%

Having dealt with the algebraic aspects, we now study how the physical action
may transform under an extended Schr\"odinger transformation. 
In particular, we are interested in the consequences for the
Schr\"odinger Ward-identities linking  the components of the 
energy-momentum tensor. We shall check
our results for free Schr\"odinger-invariant fields. 

As we have seen above, it is useful to study Schr\"odinger invariance in
a setting where the mass $\cal M$ is treated as an additional variable from the
outset. We shall therefore study two types of action. The first one
is the usual one, see e.g. \cite{Gerg02}, with $\cal M$ fixed
\BEQ
S_a = \int \!\D t\, \D \vec{r}\, {\cal L}_a \;\; , \;\;
{\cal L}_a = {\cal M} \left( \phi^{\dag}\frac{\partial\phi}{\partial t} 
- \phi\frac{\partial\phi^{\dag}}{\partial t} \right) 
+ \frac{\partial\phi^{\dag}}{\partial \vec{r}}\cdot
\frac{\partial\phi}{\partial \vec{r}}
\EEQ
The equation of motion for $S_a$ gives (\ref{1:gl:Schr}).
 The conjugate field
$\phi^{\dag}$ satisfies the same equation with $\cal M$ replaced by 
$-{\cal M}$. The `mass' $\cal M$ plays therefore the role of a quantum
number and we associate with $\phi$ a `mass' $\cal M$ and with $\phi^{\dag}$
a mass ${\cal M}^{\dag} := - {\cal M}$. 

On the other hand, treating $\cal M$ as a variable \cite{Giul96},
we use the field $\psi=\psi(\zeta,t,\vec{r})$ and consider the free action
\BEQ
S_b = \int \!\D\zeta\, \D t\, \D \vec{r}\, {\cal L}_b \;\; , \;\;
{\cal L}_b = 2\II \frac{\partial\psi}{\partial \zeta}
\frac{\partial\psi}{\partial t}
+ \left(\frac{\partial\psi}{\partial \vec{r}}\right)^2
\EEQ
with eq.~(\ref{2:gl:Schr}) as equation of motion. 

{\bf 1.} We first consider the transformation of the action $S_a$ under the
extended Schr\"odinger transformation. In view of the results of section 3,
we take $x=1/2$ as the scaling dimension of the fields $\phi,\phi^{\dag}$. 
The finite coordinate changes generated from the exponentiated generators 
$\exp \eps X_n$ and $\exp \eps Y_m$ are given by two functions 
$\beta(t)$ and $\alpha(t)$, see section 2. 
The transformation of the fields $\phi,\phi^{\dag}$ under these is
explicitly known. We find the change of the 
action $S_a \mapsto S_a^{'} = S_a +\delta S_a$, where
\BEA
\delta_X S_a &=& \int \!\D t'\, \D r'\: \frac{1}{2} {\cal M}^2 {r'}^2\, 
\left\{ \beta(t'), t'\right\} \phi'^{\dag} \phi'
\nonumber \\
\delta_Y S_a &=& \int \!\D t'\, \D r'\: {\cal M}^2 \left(
\alpha(t') - 2 r'\right) \ddot{\alpha}(t')\, \phi'^{\dag} \phi' 
\label{4:gl:Sa_trans}
\EEA
respectively. Here, 
\BEQ
\left\{ \beta(t), t\right\} = \frac{\dddot{\beta}(t)}{\dot{\beta}(t)}
-\frac{3}{2}\left(\frac{\ddot{\beta}(t)}{\dot{\beta}(t)}\right)^2
\EEQ
is the Schwarzian derivative. Consequently,
$\delta S_a=0$ if $\alpha(t)$ is at most linear in $t$ and $\beta(t)$ is
a M\"obius transformation. These transformations make up exactly the
Schr\"odinger group $\mbox{\sl Sch}(1)$ as defined in eq.~(\ref{1:gl:SCH1}),
as expected \cite{Nied72,Hage72}.   

We now discuss the Ward identities which follow from the hypothesis of
Schr\"odinger invariance. Denote by $\vec{\rho}=(t,\vec{r})$ the vector
of space-time coordinates. Consider an arbitrary infinitesimal coordinate
transformation 
$\vec{\rho}\mapsto \vec{\rho}'=\vec{\rho}+\vec{\eps}(\vec{\rho})$ such that
a field operator $\phi=\phi(t,\vec{r})$ may also pick up a phase
$\eta=\eta(t,\vec{r})$. Then the simplest possible way a local 
translation-invariant action may transform to leading order in $\eps$ is 
given by
\BEQ \label{4:gl:Wardm}
\delta S = \int\!\D t\,\D\vec{r}\: \left( T_{\mu\nu}\partial_{\mu}\eps_{\nu}
+ J_{\mu} \partial_{\mu} \eta \right)
\EEQ
which defines the energy-momentum tensor $T_{\mu\nu}$ and the 
current vector $J_{\mu}$, with $\mu,\nu=0,1,\ldots,d$. 
If we had included an extra term $J \eta$, the invariance under constant phase
shifts generated by $M_0$ would have immediately implied that $J=0$. Now,
$\delta S=0$ under translations by construction. Furthermore, scale invariance
implies the modified `trace' identity \cite{Hage72,Jack90a,Henk94}
\BEQ \label{4:gl:Wm1}
2 T_{00} + T_{11} + \ldots + T_{dd} = 0 
\EEQ
For Galilei transformations, the phase $\eta = - {\cal M}\vec{r}\cdot\vec{\eps}$
and invariance of the action $S$ implies
\BEQ \label{4:gl:Wm2}
T_{0a} + {\cal M} J_a = 0 \;\; ; \;\; a = 1, \ldots, d
\EEQ
Furthermore, using rotation invariance, one gets
$T_{ab}=T_{ba}$ with $a,b,=1,\ldots,d$. 
Now, quite analogously to conformal invariance, see \cite{diFr97,Henk99}, 
the Ward identities (\ref{4:gl:Wm1},\ref{4:gl:Wm2}) imply full Schr\"odinger 
invariance of the action. Take $d=1$ for simplicity, then $\eps_0=\eps t^2$, 
$\eps_1=\eps t r$ and $\eta=\frac{1}{2}{\cal M}r^2 \eps$. Thus for a
special Schr\"odinger transformation
\BEQ
\delta_{X_1} S = \int\!\D t\D r\: \left[ \left(2 T_{00} + T_{11} \right) t 
+\left( T_{01} +{\cal M} J_1 \right) r \right] = 0
\EEQ
Therefore, for sufficiently local interactions such that (\ref{4:gl:Wardm}) is
valid, {\em invariance under spatio-temporal translations, phase shifts, 
dilatations and Galilei transformations is enough to guarantee invariance
under the full Schr\"odinger group}. Previously, this was only proven for
vanishing masses ${\cal M}=0$ \cite{Henk94}. 

For later use, we list how $S$ should transform according to (\ref{4:gl:Wardm}) 
under the generators $X_n, Y_m$ of the extended Schr\"odinger algebra
${\cal S}_1^{\infty}$. We find
\BEQ
\delta_{X(\eps)}
S = \int \!\D t\D r\: \frac{1}{4}{\cal M} r^2\, J_0\, \dddot{\eps}
\;\; , \;\;
\delta_{Y(\eps)} S = - \int \!\D t\D r\: {\cal M} r\, J_0\, \ddot{\eps}
\EEQ
where
\BEQ \label{4:gl:eps}
X(\eps)=-\eps(t)\partial_t -\half \dot{\eps}(t) r\partial_r
-{{\cal M}\over 4} \ddot{\eps}(t) r^2 
-{x\over 2} \dot{\eps}(t)
\;\; , \;\;
Y(\eps)=-\eps(t)\partial_r-{\cal M}\dot{\eps}(t) r
\EEQ 
and by comparison with the infinitesimal transformations of $S_a$ which can
be read from eq.~(\ref{4:gl:Sa_trans}) we identify
\BEQ \label{4:gl:V0}
J_0 = 2 {\cal M} \phi^{\dag} \phi
\EEQ

To finish, we now construct $T_{\mu\nu}$ and $J_{\mu}$ explicitly from the
free-field action $S_a$. From the canonical recipe, we would obtain
\BEQ
\wit{T}_{\mu\nu} = -\delta_{\mu\nu} {\cal L}_a +
\frac{\partial{\cal L}_a}{\partial(\partial^{\mu}\phi)}\partial_{\nu}\phi
+\frac{\partial{\cal L}_a}{\partial(\partial^{\mu}\phi^{\dag})}
\partial_{\nu}\phi^{\dag} 
\;\; , \;\;
J_{\mu} = \frac{\partial{\cal L}_a}{\partial(\partial^{\mu}\phi)}\phi
-\frac{\partial{\cal L}_a}{\partial(\partial^{\mu}\phi^{\dag})}\phi^{\dag}
\EEQ
Using the equations of motion, these may be shown to satisfy 
the conservation laws 
$\partial^{\mu} \wit{T}_{\mu\nu}=\partial^{\nu}J_{\nu}=0$ 
and all Schr\"odinger 
Ward identities with the only exception of the `trace' condition 
eq.~(\ref{4:gl:Wm1}). This can be
remedied along the lines of \cite{Call70} by constructing the improved tensor
\BEQ
T_{\mu\nu} = \wit{T}_{\mu\nu} + \partial^{\lambda} B_{\lambda\mu\nu}
\EEQ
where $B$ is antisymmetric in the two first variables, 
 $B_{\lambda\mu\nu}=-B_{\mu\lambda\nu}$. If we take
$B_{a00}=\frac{d}{4}\left( \phi^{\dag}\partial_a\phi 
+\phi\partial_a\phi^{\dag}\right)$ with $a=1,\ldots,d$ 
and $B_{\lambda\mu\nu}=0$ unless $(\lambda\mu\nu)=(a00)$ (up to symmetries),
then we get
a classically conserved energy-momentum tensor, satisfying all required Ward
identities, which reads
\BEA
T_{00} &=& \frac{d{\cal M}}{2} 
\left( \phi^{\dag}\partial_t \phi -\phi\partial_t \phi^{\dag} \right) 
+\left(\frac{d}{2}-1\right) 
\partial_{\vec{r}}\phi\cdot\partial_{\vec{r}}\phi^{\dag}
\nonumber \\
T_{a0} &=& \left( 1 - \frac{d}{4}\right) 
\left( \partial_a\phi^{\dag}\partial_{t}\phi + 
\partial_a\phi\partial_{t}\phi^{\dag}\right)
-\frac{d}{4}\left(\phi^{\dag}\partial_{a,t}\phi + 
\phi\partial_{a,t}\phi^{\dag}\right) 
\nonumber \\
T_{0a} &=& {\cal M} \left( \phi^{\dag}\partial_a\phi - 
\phi\partial_a\phi^{\dag}\right)
\nonumber \\
T_{aa} &=& 2\partial_a\phi\partial_a\phi^{\dag} -
\partial_{\vec{r}}\phi\cdot\partial_{\vec{r}}\phi^{\dag} -
{\cal M} \left( \phi^{\dag}\partial_t\phi - 
\phi\partial_t\phi^{\dag}\right)
\nonumber \\
T_{ab} &=& \partial_a\phi\partial_b\phi^{\dag}+ 
\partial_b\phi\partial_a\phi^{\dag}
\EEA
The current $J_{\mu}$ need not be improved and we have
\BEQ
J_0 = 2{\cal M} \phi^{\dag}\phi \;\; , \;\;
J_a = \phi\partial_a\phi^{\dag} - \phi^{\dag}\partial_a\phi
\EEQ
In particular, we recover (\ref{4:gl:V0}). 
For a physical interpretation, it is better to divide 
${\cal L}_a$ by $2{\cal M}$. We then recover the usual interpretation of 
$J_0$ as a probability density, $J_a$ as a probability current, $T_{00}$
as an energy density, $T_{0a}$ as a momentum density and $T_{a0}$ and $T_{ab}$
as energy and momentum currents, respectively. 

Notice that $(-2{\cal M})^{-1}T_{00}$ coincides in $d=2$ dimensions
and with $t$ replaced by $z$, with the energy-momentum 
tensor $T(z)$ of a complex fermionic field, see \cite[p. 147]{diFr97}.

{\bf 2.} Similarly, we now treat the transformation of the 
free-field action $S_b$
under the $3D$ conformal group $\mbox{\sl Conf\/}(3)$, whose Lie algebra 
generators are listed in section 3. Recalling the transformation of the field 
$\psi$ from section 2, and setting $x=1/2$, the action changes into
$S_b \mapsto S_b^{'} = S_b + \delta S_b$ where
\BEA
\delta_X S_b &=& \int \!\D\zeta'\,\D t'\, \D r'\: \frac{1}{2} {r'}^2\, 
\left\{ \beta(t'), t'\right\} 
\left(\frac{\partial\psi'}{\partial \zeta'}\right)^2
\nonumber \\
\delta_Y S_b &=& \int \!\D\zeta'\,\D t'\, \D r'\:  \left(
\alpha(t') - 2 r'\right) \ddot{\alpha}(t') 
\left(\frac{\partial\psi'}{\partial \zeta'}\right)^2
\label{4:gl:Sb_trans}
\EEA
It is also esay to see that $\delta S=0$ under the other generators of
${\mathfrak{conf}}_3$. This means that $S_b$ is
invariant under the $3D$ conformal group, in agreement with 
the conclusion drawn in section 3 from the infinitesimal transformations. 

As before, we now discuss the Ward identities. To straighten notation, we
construct a vector $\vec{\xi}$ with components
\BEQ
\xi_{-1} = \zeta \;\; , \;\; \xi_0=t \;\; , \;\; \xi_1 = r_1 \;\;
,\;\; \ldots\;\; ,\;\; \xi_d = r_d
\EEQ
and write the derivatives $\partial^{\mu}\psi=\partial\psi/\partial\xi_{\mu}$. 
Under an infinitesimal transformation 
$\vec{\xi}\mapsto\vec{\xi}'=\vec{\xi}+\vec{\eps}(\vec{\xi})$, the 
action is assumed to transform to leading order as
\BEQ \label{4:gl:Wardz}
\delta S = \int \!\D\zeta\D t\D\vec{r}\: T_{\mu}^{\nu} \partial^{\mu}
\eps_{\nu}
\EEQ
and we proceed to write down the Ward identities, restricting to $d=1$ for
simplicity.  Again, $S$ is 
translation-invariant par construction. Dilatation invariance implies
\BEQ \label{4:gl:Wz1}
2 T_{0}^{0} + T_{1}^{1} = 0
\EEQ
Invariance of $S$ under the three generators $N$, $Y_{1/2}$ and 
${V}_{-}$ coming from the $3D$ conformal rotations (see section 3)
yields the following Ward identities, respectively
\BEQ \label{4:gl:Wz2}
T_{-1}^{-1} - T_{0}^{0} = 0 \;\; , \;\;
T_{0}^{1} - \II T_{1}^{-1} = 0 \;\; , \;\;
T_{-1}^{1} -\II T_{1}^{0} = 0 
\EEQ
and it follows that $T$ has 5 independent components. Next, we consider 
the three remaining generators $X_1$, $W$ and $V_+$ which come
from the $3D$ special conformal transformations. The components of 
$\eps_{\nu}$ are easily read from the generators and we have
\BEA
\delta_{X_{1}}S &=& -\eps \int\!\D\zeta\D t\D r\: 
\left[ \left(2T_{0}^{0}+T_{1}^{1}\right) t + 
\left(T_{0}^{1}-\II T_{1}^{-1}\right) r \right] = 0
\nonumber \\
\delta_{W}S &=& -\eps \int\!\D\zeta\D t\D r\: 
\left[ \left(2T_{-1}^{-1}+T_{1}^{1}\right) \zeta + 
\left(T_{-1}^{1}-\II T_{1}^{0}\right) r \right] = 0
\nonumber \\
\delta_{V_+}S &=& -\eps \int\!\D\zeta\D t\D r\: 
\left[ \left(T_{-1}^{-1}+T_{0}^{0}+T_{1}^{1}\right) r + 
\left(T_{0}^{1}-\II T_{1}^{-1}\right) \II\zeta + 
\left(T_{-1}^{1}-\II T_{1}^{0}\right) \II t \right] = 0
\label{4:gl:WardXWV}
\EEA
This merely translates the well-known result of 
conformal invariance, namely 
that translation, rotation and scale invariance 
imply full conformal invariance
\cite{Call70,diFr97,Henk99}, to the formulation at hand. Furthermore,
considering the transformation of $S$ under the 
infinitesimal action of the generators
$X$ and $Y$ of the extended Schr\"odinger algebra, we read off $\eps_{\nu}$
from the Fourier transform of (\ref{4:gl:eps}) and find
\BEQ 
\delta_{X(\eps)} S = \frac{\II}{4}\int\!\D\zeta\D t\D r\:
r^2\,T_{0}^{-1}\,\dddot{\eps}
\;\; , \;\;
\delta_{Y(\eps)} S = {\II}\int\!\D\zeta\D t\D r\:r\,T_{0}^{-1}\,\ddot{\eps}
\EEQ
As before, we can compare this with the infinitesimal form of the
transformation of the free-field action $S_b$ in eq.~(\ref{4:gl:Sb_trans})
and identify
\BEQ \label{4:gl:T0-1}
T_{0}^{-1} = 2\II \left( \frac{\partial\psi}{\partial\zeta}\right)^2
\EEQ
which is the analogue of eq.~(\ref{4:gl:V0}). 

Therefore, the current $J_0$ or the component $T_{0}^{-1}$, respectively, 
generate the change of the action under an infinitesimal extended Schr\"odinger
transformation. We point out that in $2D$ conformally invariant classical
field theories, no such term is present. Namely, the conformal generators
$\ell_n = - z^{n+1}\partial_z$ and $\overline{\ell}_n = - \overline{z}^{n+1}
\partial_{\overline{z}}$ with $n\in\mathbb{Z}$ can be rewritten as
$X_n = \ell_n + \overline{\ell}_n$ and $Y_n = -\II(\ell_n - \overline{\ell}_n)$
where the `time' $t$ and `space' $r$ were introduced from the complex
coordinates $z,\overline{z}$ via $z=t+\II r, \overline{z}=t-\II r$. The
projective conformal Ward identities $T_{00}+T_{11}=0$ and $T_{01}=T_{10}$
follow as usual. Discarding any conformal anomalies, these identities 
are sufficient to show that $\delta_{X_n} S = \delta_{Y_n} S =0$ 
for all $n\in\mathbb{Z}$.

We finish by constructing the energy-momentum tensor explicitly. 
The canonical energy-momentum tensor is given by 
\BEQ
\wit{T}_{\mu}^{\nu}=-\delta_{\mu}^{\nu} {\cal L}_b +{\partial {\cal L}_b
\over \partial (\partial^{\mu}\psi)} \partial^{\nu} \psi.
\EEQ
and may be written in a matrix form (here for the $d=1$ case, where $\nu$ 
labels the columns and we write 
$\psi_{\zeta}=\partial\psi/\partial\zeta,\ldots$)
\BEQ
\wit{T}_{\mu}^{\nu}=\left(\begin{array}{ccc} 
-\psi_r^2 & 2\II\psi_t^2 & 2\II\psi_r \psi_t \\
2\II\psi_{\zeta}^2 & -\psi_r^2 & 2\II\psi_{\zeta}\psi_r \\
2\psi_r \psi_{\zeta} & 2\psi_r \psi_t & \psi_r^2 -2\II\psi_{\zeta}\psi_t
\end{array}\right).
\EEQ
It reproduces (\ref{4:gl:T0-1}), is classically conserved and satisfies 
all Ward identities  (\ref{4:gl:Wz2}) coming from the $3D$ conformal group, 
but not scaling identity eq.~(\ref{4:gl:Wz1}). To correct this,
define the improved tensor \cite{Call70}
\BEQ
T_{\mu}^{\nu}=\wit{T}_{\mu}^{\nu}+\partial^{\rho} B_{\rho \mu}^{\nu},
\EEQ
with $B_{\rho \mu}^{\nu}$  antisymmetric in $\rho$ in $\mu$, which has
the same divergence as $\wit{T}$. Choosing 
\BEQ
B_{-1 0}^{\nu}=\left(\begin{array}{c} -{\II\over 2} \psi\psi_{\zeta} \\
{\II\over 2} \psi\psi_t \\ 0 \end{array}\right),
B_{1 0}^{\nu}=\left(\begin{array}{c} 0 \\ {1\over 2}\psi\psi_r \\-{\II\over 2}
\psi\psi_{\zeta} \end{array}\right),
B_{1 -1}^{\nu}=\left(\begin{array}{c} {1\over 2} \psi\psi_r \\ 0\\-{\II\over 2}
\psi\psi_t \end{array}\right),
\EEQ
the improved energy-momentum tensor reads (where the equations of motion
were used)  
\BEQ
T_{\mu}^{\nu} = \frac{1}{2}\left(\begin{array}{ccc}
{\II}\psi_{\zeta}\psi_t-\II\psi\psi_{\zeta t} -\psi_r^2 &
  {3}\II\psi_t^2-{\II}\psi\psi_{tt} & 
  {\II}[3\psi_t \psi_r-\psi\psi_{tr}] \\
3{\II} \psi_{\zeta}^2 -{\II}\psi\psi_{\zeta\zeta} & 
{\II}\psi_{\zeta}\psi_t -\II\psi\psi_{\zeta t} -\psi_r^2
   & {\II}[3 \psi_{\zeta}\psi_r -\psi\psi_{\zeta r}] \\
{3}\psi_{\zeta}\psi_r - \psi\psi_{\zeta r} & 
  {3}\psi_t \psi_r-\psi\psi_{t r} &
-2[\II\psi_{\zeta}\psi_t -\II\psi\psi_{\zeta t} -\psi_r^2]  
\end{array} \right) .
\EEQ
which manifestly satisfies the conditions eqs.~(\ref{4:gl:Wz1},\ref{4:gl:Wz2})
and is conserved. Of course, this is just a
particular case of the construction of the Belinfante tensor in $3D$ conformal 
theory (see \cite{Call70,diFr97}). As pointed out long ago \cite{Call70}, the
consideration of the improved energy-momentum tensor in Poincar\'e-invariant
interacting theories is of particular interest, since satisfying the
conformal Ward identities implies that the elements of $T_{\mu}^{\nu}$ remain
finite in the limit of a large cut-off for renormalizable interactions. 
Their result should translate, through the inclusion of ${\mathfrak
{sch}}_d$ into ${\mathfrak{conf}}_{d+2}$  described in section 3, to 
non-relativistic interacting theories and one should be able to avoid this
way difficulties \cite{Hage72a} with the finiteness of the elements of 
$T_{\mu\nu}$ which may arise in Galilei-invariant field theories with 
fixed masses.

{\bf 3.} Having studied so far the full 
conformal algebra $(\mathfrak{conf}_3)_{\C}$,
we now consider its subalgebras. In particular, we inquire about the status
of the Ward identities for the subalgebras $\mathfrak{sch}_1$ and
$\mathfrak{age}_1$. 

We consider first the free-field action $S_b$, formulated in terms of scaling
operators $\psi=\psi(\zeta,t,r)$. From eq.~(\ref{4:gl:WardXWV}) it is clear 
that the Ward identities coming from dilatation and Galilei invariance are
sufficient to prove also special Schr\"odinger invariance, i.e.
$\delta_{X_1} S=0$. The question raised is thus settled for $\mathfrak{sch}_1$. 

A new aspect arises, however, for  $\mathfrak{age}_1$. In that case, time
translation invariance is no longer assumed, but all elements of 
$\mathfrak{age}_1$ keep the line $t=0$ invariant. Consequently, the 
transformation (\ref{4:gl:Wardz})  of the action under infinitesimal 
transformations might be generalized to
\BEQ \label{4:gl:29}
\delta S = \int \!\D\zeta\D t\D\vec{r}\: T_{\mu}^{\nu} \partial^{\mu}
\eps_{\nu} + \int_{(t=0)} \!\!\!\D\zeta\D\vec{r}\: U^{\nu}\eps_{\nu}
\EEQ
where the second integral is restricted to the `boundary' $t=0$. 

Then, and now specializing to $d=1$, 
translation invariance in $r$ and in $\zeta$ yields $U^{1}=U^{-1}=0$.
Although $U^0$ is not fixed, it does not contribute to $\delta S$, since
$\eps_0=0$ at $t=0$ for all elements of $\mathfrak{age}_1$. From dilatation and
Galilei invariance, the Ward identities $T_0^0 + \frac{1}{2} T_1^1=0$ and
$T_0^{1}-\II T_1^{-1}=0$ follow and consequently for a special 
Schr\"odinger transformation generated by $X_{1}$, we have
\BEQ
\delta_{X_1} S = -\eps \int\!\D\zeta\D t\D r\: \left[ 
\left( 2 T_0^0 + T_1^1\right) t + \left( T_0^1 -\II T_1^{-1}\right)r\right]
+\frac{\II\eps}{2} \int_{(t=0)} \!\D\zeta\D r\: r^2 U^{-1} = 0
\EEQ
This means that the validity of special Schr\"odinger invariance mainly
depends on having a $z=2$ scale invariance and Galilei invariance, while
time-translation invariance is not really required. 

Similarly, if we had chosen instead to work with a fixed mass ${\cal M}$, the
transformation (\ref{4:gl:Wardm}) of the action should be generalized 
as follows
\BEQ \label{4:gl:31}
\delta S = \int\!\D t\,\D\vec{r}\: \left( T_{\mu\nu}\partial_{\mu}\eps_{\nu}
+ J_{\mu} \partial_{\mu} \eta \right) 
+\int_{(t=0)} \!\!\!\D\vec{r}\: \left( U_{\nu} \eps_{\nu} + V \eta \right)
\EEQ
and it is now straightforward to see that again special Schr\"odinger invariance
$\delta_{X_1}S=0$ follows. Note that it follows in particular form phase-shift 
invariance that $V$ should be $0$.

Our result is that special Schr\"odinger invariance holds as a consequence of
scale and Galilei invariance, even in the absence of time-translation 
invariance, provided only that the dynamics is `local' in the 
sense of eqs.~(\ref{4:gl:29}) or (\ref{4:gl:31}). It is well-known that
ageing ferromagnets undergoing phase-ordering 
kinetics after being quenched to a fixed temperature $T$ below criticality 
is scale invariant with a dynamical exponent $z=2$ \cite{Bray94}. If we
accept that these systems are also Galilei-invariant, we can conclude
that these coarsening systems must also be invariant under special
Schr\"odinger transformations. This explains
{\em why} special Schr\"odinger invariance could be successfully
tested in these systems, see \cite{Henk01,Godr02,Pico02,Henk02a,Henk02b}. 

%%%%%%%%%%%%%%%%%%%%%%%%%%%%%%%%%%%%%%%%%%%%%%%%%%%%%%%%%%%%%%%%%%%%%%%%%%%%%%%%
\section{Response functions}
%%%%%%%%%%%%%%%%%%%%%%%%%%%%%%%%%%%%%%%%%%%%%%%%%%%%%%%%%%%%%%%%%%%%%%%%%%%%%%%%

Having described the relationship between conformal and Schr\"odinger
transformations, we now discuss the consequences for the two- and three-point
functions. Let $\Psi_a(\vec{\xi}_a)=\psi_a(\zeta_a,t_a,\vec{r}_a)$
be a scalar and quasi-primary conformal scaling operators with scaling 
dimension $x_a$. We can always take the $\Psi_a$ to be real. 
It is well-known that \cite{Poly70}
\BEA
\lefteqn{
\left\langle \psi_1(\zeta_1,t_1,\vec{r}_1)
\psi_2(\zeta_2,t_2,\vec{r}_2)\right\rangle =
\left\langle \Psi_1(\vec{\xi}_1) \Psi_2(\vec{\xi}_2) \right\rangle =
\Psi_0 \delta_{x_1,x_2}\, |\vec{\xi}_1-\vec{\xi}_2|^{-2x_1} 
} 
\nonumber \\
&=& \psi_0 \delta_{x_1,x_2}\, (t_1-t_2)^{-x_1} 
\left( \zeta_1-\zeta_2+{\II\over 2} {(\vec{r}_1-\vec{r}_2)^2\over
t_1-t_2}\right)^{-x_1}
\label{5:gl:1} 
\EEA
where $\psi_0 = 4^{-x_1}\Psi_0$ and $\Psi_0$ is a normalization constant. 
In order to understand the physical meaning of the result (\ref{5:gl:1}), 
we rewrite it in terms of scaling operators $\phi_a(t,r)$
with fixed mass ${\cal M}_a\ge 0$, using eq.~(\ref{2:gl:phipsi}). 
As shown in appendix B, we find, provided $x_1>0$
\BEA
\left\langle \phi_1(t_1,\vec{r}_1)
\phi_2^*(t_2,\vec{r}_2)\right\rangle &=& {1\over 2\pi}
\int_{\R^2} \!\D\zeta_1 \D\zeta_2\: 
e^{-\II{\cal M}_1\zeta_1+\II{\cal M}_2\zeta_2}
\left\langle \psi_1(\zeta_1,t_1,\vec{r}_1)
\psi_2(\zeta_2,t_2,\vec{r}_2)\right\rangle 
\label{5:gl:2} \\
&=& \phi_0\, \delta_{x_1,x_2}\, 
\delta({\cal M}_1-{\cal M}_2)\, {\cal M}_1^{1-x_1}\: 
\Theta(t_1-t_2)  (t_1-t_2)^{-x_1} \exp\left(-\frac{{\cal M}_1}{2}
\frac{(\vec{r}_1-\vec{r}_2)^2}{t_1-t_2}\right)
\nonumber 
\EEA
where $\phi_0$ is again a normalization constant (proportional to $\psi_0$) 
and $\Theta$ is the Heaviside
function. The functional form of $\langle\phi\phi^*\rangle$ 
had been derived before from the requirement that it transform covariantly 
under the realization of the Schr\"odinger group {\sl Sch}($d$) with fixed
masses \cite{Henk94}. We stress that the causal
prefactor $\Theta(t_1-t_2)$ is {\it not} a consequence of covariance under that
realization, but rather had to be put in by hand \cite{Henk94} 
to comply with the requirement that 
$\langle \phi_1 \phi_2^*\rangle$ should decay to zero for large distances 
$r=|\vec{r}_1-\vec{r}_2|$. Furthermore, 
causality as implied in (\ref{5:gl:2}) fits perfectly with the 
interpretation of $\langle \phi \phi^*\rangle=\langle \phi \wit{\phi}\rangle$ 
as response function in the context of Martin-Siggia-Rose theory \cite{Mart73} 
and suggests the identification of the complex conjugate scaling 
operator $\phi^*$ with the response operator $\wit{\phi}$, 
conjugate to the order parameter scaling operator $\phi$. 

An analogous result holds for three-point functions. Recall the well-known
result of conformal invariance \cite{Poly70}
\BEQ \label{5:gl:3}
\left\langle \Psi_1(\vec{\xi}_1) \Psi_2(\vec{\xi}_2) \Psi_3(\vec{\xi}_3)
\right\rangle
= C_{123}\, |\vec{\xi}_1-\vec{\xi}_2|^{-x_{12,3}}\:
|\vec{\xi}_2-\vec{\xi}_3|^{-x_{23,1}}\:|\vec{\xi}_3-\vec{\xi}_1|^{-x_{31,2}} 
\EEQ
where $x_{ab,c}:=x_a+x_b-x_c$ and $C_{123}$ is an operator product expansion
(OPE) coefficient. As shown in appendix B we find, provided
$x_1>0$ and $x_2>0$
\BEA
\lefteqn{
\left\langle \phi_1(t_1,\vec{r}_1)\phi_2(t_2,\vec{r}_2)
\phi_3^*(t_3,\vec{r}_3)\right\rangle
= {\cal C}_{12,3}\, \delta({\cal M}_1+{\cal M}_2 - {\cal M}_3)\: 
}
\nonumber \\
&\times& \Theta(t_1-t_3)\, \Theta(t_2-t_3)\:  
(t_1-t_2)^{-x_{12,3}/2}\,(t_1-t_3)^{-x_{13,2}/2}\,
(t_2-t_3)^{-x_{23,1}/2} 
\label{5:gl:3P} \\ 
&\times& \exp\left[ -\frac{{\cal M}_1}{2} 
\frac{(\vec{r}_1-\vec{r}_3)^2}{t_1-t_3}
-\frac{{\cal M}_2}{2} \frac{(\vec{r}_2-\vec{r}_3)^2}{t_2-t_3} \right]
\Phi_{12,3} \left( \frac{1}{2} 
\frac{[(\vec{r}_1-\vec{r}_3)(t_2-t_3)-
(\vec{r}_2-\vec{r}_3)(t_1-t_3)]^2}{(t_1-t_2)(t_2-t_3)(t_1-t_3)}
\right)
\nonumber 
\EEA
where ${\cal C}_{12,3}$ is a constant related to the OPE coefficient
$C_{123}$ and $\Phi_{12,3}(v)$ is a scaling function (see eq.~(\ref{B:eq:11})
for an integral representation). As we had seen before for
the two-point function, the functional form of this result is in complete
agreement with the one found previously from Schr\"odinger invariance alone
for scaling operators $\phi_a$ with fixed masses 
${\cal M}_a\geq 0$ \cite{Henk94}. However, the causality conditions
$t_1>t_3$ and $t_2>t_3$ expected for a response function are now
automatically satisfied and need no longer be put in by hand. The conditions
on the masses in (\ref{5:gl:2},\ref{5:gl:3P}) 
are nothing but examples of the standard Bargmann superselection rules. 

In summary, we have seen how to reconstruct the physically more appealing
expectation values $\langle\phi\phi^*\rangle$ and
$\langle\phi\phi\phi^*\rangle$ from their conformal
forms as given by $\langle\psi\psi\rangle$, $\langle\psi\psi\psi\rangle$,
where we have used right away the forms (\ref{5:gl:1},\ref{5:gl:3}) 
which follow from covariance under the full 
conformal algebra $(\mathfrak{conf}_3)_{\C}$. 
We now ask to what extent the form
of the two-point function is already determined by one of the subalgebras
as obtained in section 3. 

In order to do so, we shall begin by deriving the constraints on the
two-point function (for simplicity in $d=1$ dimensions)
\BEQ
F = \left\langle \psi_1(\zeta_1,t_1,r_1)\psi_2(\zeta_2,t_2,r_2)\right\rangle
\EEQ
coming from its invariance under the minimal parabolic subalgebra 
$\wit{\mathfrak{age}}_1=\{X_{0,1}, Y_{\pm 1/2}, M_0, N\}$. It is convenient
to introduce the variables $u=t_1-t_2$ and $v=t_1/t_2$. From space and phase
translation invariance, we have $F=F(\zeta,u,v,r)$, where
$\zeta=\zeta_1-\zeta_2$ and $r=r_1-r_2$. Next, the covariance conditions
yield the following equations
\BEA
X_0 F &=& \left( - u \partial_u - \frac{1}{2} r \partial_r - \frac{x_1+x_2}{2}
\right) F = 0 
\nonumber \\
Y_{1/2} F &=& \left( - u \partial_r + \II r \partial_{\zeta} \right) F = 0 
\nonumber \\
N F &=& \left( -u \partial_u + \zeta \partial_{\zeta} \right) F = 0 
\label{5:gl:XYNXcov} \\
X_1 F &=& \left( - u^2 \partial_u - u v \partial_v - u r \partial_r
 +\frac{\II}{2} r^2
\partial_{\zeta} - u x_1 \right) F = 0
\nonumber 
\EEA
Using the first and second of these again, the last condition
$X_1F=0$ simplifies into
\BEQ \label{5:gl:X1cov}
\left( - v \partial_v + \frac{x_2-x_1}{2} \right) F = 0
\EEQ
Therefore, we have the factorization $F=F_0(v) F_1(\zeta,u,r)$, where
$F_1$ satisfies the same conditions found in the time-translation invariant
case. We easily find
\BEQ \label{5:gl:psiuv}
\left\langle \psi_1(\zeta_1,t_1,r_1)
\psi_2(\zeta_2,t_2,r_2)\right\rangle 
= \psi_0 \left( \frac{t_1}{t_2}\right)^{(x_2-x_1)/2}\, 
\left(t_1-t_2\right)^{-(x_1+x_2)/2}\, 
\left( \zeta_1-\zeta_2+{\II\over 2} 
{(r_1-r_2)^2\over t_1-t_2}\right)^{-(x_1+x_2)/2}
\EEQ
with some normalization constant $\psi_0$. Comparing with  the conformal 
result eq.~(\ref{5:gl:1}), we observe the absence of the constraint $x_1=x_2$
on the scaling dimensions and the presence of a factor  which explicitly
breaks time-translation invariance. 

If we had merely required invariance under a proper subalgebra of 
the minimal parabolic subalgebra $\wit{\mathfrak{age}}_1$, we would not have had
sufficiently many conditions in order to fix the two-point function completely.
However, if we restrict ourselves to $\mathfrak{age}_1$ by leaving out only the 
generator $N$, although the two-point function 
$\langle\psi\psi\rangle=t^{-x_1}E(\zeta+\II r^2/2t)$ then contains the 
undetermined scaling function $E$, the form of the response function
$\langle\phi\phi^*\rangle=\phi_0 t^{-x_1}\exp(-r^2/(2{\cal M}_1 t))$
is still completely fixed, up to a mass-dependent normalization constant
$\phi_0=\phi_0({\cal M}_1)$. 

Next, we consider the extension of $\wit{\mathfrak{age}}_1$ to one of the 
two maximal parabolic subalgebras of ${\mathfrak{conf}}_3$ studied in
section 3 and in appendix~C.
First, we consider $\wit{\mathfrak{sch}}_1$ 
by adding the generator $X_{-1}$ of time translations. 
This simply enforces $x_1=x_2$ and
we recover the conformal result (\ref{5:gl:1}). Second, we consider
$\wit{\mathfrak{alt}}_1$ by adding the generator $V_+$. We obtain
\BEA
V_+ F &=& \left( - 2 u r \partial_u - 2 v r\partial_v - 
2 \zeta r \partial_{\zeta}
-(r^2+2\II\zeta u)\partial_r - 2 r x_1 - 2t_2\left( r\partial_u +\II \zeta
 \partial_r \right)
\right) F \nonumber \\
&=& -2u(v-1)^{-1}\left( r\partial_u +\II \zeta \partial_r\right) F 
\EEA
and where in the second line the covariance conditions 
eqs.~(\ref{5:gl:XYNXcov},\ref{5:gl:X1cov}) were used. From (\ref{5:gl:psiuv})
it can be seen that indeed $V_+F=0$ and the two-point function transforms
covariantly under $\wit{\mathfrak{alt}}_1$. In this case, there is no
constraint on $x_1$ and $x_2$.

The case of an arbitrary space dimension $d$ is treated in the same way
by simply replacing the $r_a$ by $\vec{r}_a$. 
In conclusion, we have found two distinct forms of the two-point function, 
namely (\ref{5:gl:1}) and (\ref{5:gl:psiuv}). 
The form of the two-point function is completely fixed if the symmetry
algebra contains $\wit{\mathfrak{age}}$. 
If in addition the dynamic symmetry algebra contains the extended 
Schr\"odinger algebra $\wit{\mathfrak{sch}}$, 
the two-point function $\langle\psi\psi\rangle$ 
reduces to the form (\ref{5:gl:1}) as obtained from full conformal invariance
or, alternatively, the causal form $\langle\phi\phi^*\rangle$ of
eq.~(\ref{5:gl:2}). On the other hand, if time-translation invariance is
absent, the algebra of dynamical symmetries may be given by
$\wit{\mathfrak{alt}}$ or $\wit{\mathfrak{age}}$ and the two-point function
is given by (\ref{5:gl:psiuv}). The causal form with fixed masses is then
given by, provided $x_1+x_2>0$ 
\BEA
\lefteqn{ 
\left\langle \phi_1(t_1,\vec{r}_1)\phi_2^*(t_2,\vec{r}_2)\right\rangle = 
\phi_0\, \delta({\cal M}_1-{\cal M}_2)\, {\cal M}_1^{1-(x_1+x_2)/2}\: 
}
\nonumber \\
&\times& \Theta(t_1-t_2)  \left(\frac{t_1}{t_2}\right)^{(x_2-x_1)/2} 
(t_1-t_2)^{-(x_1+x_2)/2} \exp\left(-\frac{{\cal M}_1}{2}
\frac{(\vec{r}_1-\vec{r}_2)^2}{t_1-t_2}\right)
\label{5:gl:2A}
\EEA
This form for the response function $R(t,s;\vec{r})
=\langle\phi(t;\vec{r})\wit{\phi}(s;\vec{0})\rangle=
\langle\phi(t;\vec{r})\phi^*(s;\vec{0})\rangle$, including the
causality condition $t>s$, has been confirmed
in ageing phenomena in several models of simple ferromagnets quenched to a 
temperature $T<T_c$ below criticality, in particular the
$2D$ and $3D$ Glauber-Ising model \cite{Henk01,Henk02a,Henk02b}, the
spherical model with a non-conserved order parameter 
\cite{Cala02,Cann01,Godr00b,Henk01,Pico02} and, last but not least, the
free random walk \cite{Cugl94}.

%%%%%%%%%%%%%%%%%%%%%%%%%%%%%%%%%%%%%%%%%%%%%%%%%%%%%%%%%%%%%%%%%%%%%%%%%%%%%%%%
\section{Conclusions}
%%%%%%%%%%%%%%%%%%%%%%%%%%%%%%%%%%%%%%%%%%%%%%%%%%%%%%%%%%%%%%%%%%%%%%%%%%%%%%%%

Our study of Schr\"odinger invariance as a dynamical space-time symmetry was
motivated by the known explicit confirmation of some of its consequences in a 
few specific and non-trivial models. In order to understand better the origin 
of such a symmetry, a useful starting point is the analysis of the associated
classical free-field theory of which the Schr\"odinger equation is the
Euler-Lagrange equation of motion. Going through this exercise, 
it became apparent to us that the mass $\cal M$ in this 
equation should be considered as a dynamical
variable on the same level as space and time coordinates. This leads us to
the following results: 
\begin{enumerate}
\item the usual projective representation of the Schr\"odinger group becomes
a true representation, via conjugation by Fourier transformation with
respect to $\cal M$.
\item  a new relation between the Schr\"odinger Lie algebra $\mathfrak{sch}_d$ 
and the complexified conformal Lie algebra $(\mathfrak{conf}_{d+2})_{\C}$ is
found, namely
\BEQ
\mathfrak{sch}_d \subset (\mathfrak{conf}_{d+2})_{\C}
\EEQ
Some subalgebras of 
$(\mathfrak{conf}_{d+2})_{\C}$ closely related to parabolic subalgebras may
play a r\^ole in physical applications, notably to ageing phenomena in
spin systems. We leave the elaboration of this to future work. 

We also reconsidered an old claim \cite{Baru73} that $\mathfrak{sch}_d$ could
be obtained as a non-relativistic limit from a conformal Lie algebra. 
If the mass $\cal M$ is treated as a dynamic variable from the
beginning, we find in the non-relativistic limit $c\to\infty$ instead 
$(\mathfrak{conf}_{3})_{\C}\to\wit{\mathfrak{alt}}_1\ne\mathfrak{sch}_1$. 
\item the Ward identities which express the invariance of an action under 
conformal transformations, or merely under the Schr\"odinger subalgebra
$\mathfrak{sch}_d$ or else $\mathfrak{age}$, are derived from the `locality
assumption' eqs.~(\ref{4:gl:29}) or (\ref{4:gl:31}). An important open
question is the extension of $\mathfrak{sch}_d$ to an  
infinite-dimensional algebra - possibly ${\cal S}_1^{\infty}$ or
natural extensions thereof to $d$
spatial dimensions - such as to include those 
terms which would describe the contributions of anomalies coming from quantum 
fluctuations. In order to prepare such a study, 
we explicitly constructed the conserved
energy-momentum tensor (and also the probability current) which satisfy these
Schr\"odinger (or conformal) Ward identities. 
Further work along these lines is in progress. 

When applying these considerations to ageing, we must take into account that
time-translation invariance does no longer hold. This can be dealt with 
by allowing for a `boundary term' in the action and coming from the
$t=0$ initial conditions. Schematically, we have  
\BEQ
\left. \begin{array}{c}
\mbox{\rm spatial translation invariance} \\
\mbox{\rm phase shift invariance}\\
\mbox{\rm Galilei invariance} \\
\mbox{\rm scale invariance with $z=2$} \\
\mbox{\rm locality}
\end{array} \right\} \Longrightarrow 
\mbox{\rm special Schr\"odinger invariance} 
\EEQ
where locality is understood in the sense of 
eqs.~(\ref{4:gl:29},\ref{4:gl:31}). 
We emphasize that while time-translation invariance is not really needed,
Galilei invariance is a necessary condition for having an invariance under
the action of the special Schr\"odinger transformation generated by $X_1$. 

We have also seen that the classical free-field action is not invariant under 
the algebra ${\cal S}_1^{\infty}$, in contrast with the situation found for 
two-dimensional conformal invariance. 
\item tests of the predictions of Schr\"odinger invariance can 
be carried out by considering the following response functions
\BEA
R_2(t_1,s)&=&\frac{\delta \langle\phi(t_1)\rangle}{\delta h(s)} = 
\left\langle \phi(t_1) \wit{\phi}(s)\right\rangle = 
\left\langle \phi(t_1) {\phi}^*(s)\right\rangle 
\nonumber \\ 
R_3(t_1,t_2,s)&=&\frac{\delta \langle\phi(t_1)\phi(t_2)\rangle}{\delta h(s)} = 
\left\langle \phi(t_1)\phi(t_2) \wit{\phi}(s)\right\rangle = 
\left\langle \phi(t_1)\phi(t_2) {\phi}^*(s)\right\rangle 
\EEA
where $\wit{\phi}$ is the Martin-Siggia-Rose response operator associated with 
the order parameter scaling operator $\phi$ and $h$ is the conjugate magnetic 
field. Here $t_1, t_2$ are observation times and $s$ is a waiting time. 
Our results (\ref{5:gl:2},\ref{5:gl:3P},\ref{5:gl:2A}) suggest the 
identification $\wit{\phi}=\phi^*$ of the response operator $\wit{\phi}$ 
with the `complex conjugate'
$\phi^*$ in the formalism at hand. In particular, the causality conditions
$t_1>s$ and $t_2>s$ required for an interpretation of $R_2$ and $R_3$ as
response function were derived in a model-independent way. 

In several spin systems undergoing ageing, the resulting scaling form 
(\ref{5:gl:2A}) of the two-point response function $R_2$ has
been fully confirmed, see
\cite{Henk01,Godr02,Pico02,Henk02,Henk02a,Henk02b}. However, we are not aware
of any tests of (\ref{5:gl:2},\ref{5:gl:3P}) in {\em equilibrium} critical 
dynamics. It appears 
significant that covariance under at least the minimal parabolic subalgebra of 
the conformal algebra is required in order to fix the two-point function 
completely. Tests of the three-point response $R_3$ would be very interesting 
and might provide an answer to the open question which of the two parabolic
subalgebras $\wit{\mathfrak{alt}}$ or $\wit{\mathfrak{age}}$ is the
relevant one for the description of ageing phenomena. This will also require
the explicit inclusion of the effects of the initial conditions into the 
analysis, where work is in progress \cite{Pico03}. 
\end{enumerate}

\newpage 

%%%%%%%%%%%%%%%%%%%%%%%%%%%%%%%%%%%%%%%%%%%%%%%%%%%%%%%%%%%%%%%%%%%%%%%%%%%%%%%%
\appsection{A}{On the non-relativistic limit}
%%%%%%%%%%%%%%%%%%%%%%%%%%%%%%%%%%%%%%%%%%%%%%%%%%%%%%%%%%%%%%%%%%%%%%%%%%%%%%%%

We briefly describe the non-relativistic limit of the conformal Lie algebra
$\mathfrak{conf}_{d}$ and the relation to the Schr\"odinger Lie algebra
$\mathfrak{sch}_d$. This was discussed by Barut \cite{Baru73} long ago. 
He started from the massive Klein-Gordon equation
\BEQ \label{A:gl:1}
\left( \frac{1}{c^2}\frac{\partial^2}{\partial t^2} +
\frac{\partial}{\partial \vec{r}}\cdot\frac{\partial}{\partial \vec{r}} 
-{\cal M}^2 c^2 \right) \vph_{\cal M}(t,\vec{r}) = 0
\EEQ
where $c$ is the speed of light. This equation is invariant under the
conformal group in $(d+1)$ dimensions, {\em provided} the mass $\cal M$ is
transformed as well \cite{Baru73}. After the substitution 
\BEQ \label{A:gl:subs} 
\partial_t \mapsto {\cal M}c + \frac{1}{c}\partial_t
\EEQ
the non-relativistic limit $c\to\infty$ reduces (\ref{A:gl:1}) to the
Schr\"odinger equation. However, the $c\to\infty$ limit of the 
conformal generators (\ref{3:gl:konfG}) in $d+1$ dimensions (and with 
$\xi_0=ct$ and $\xi_a=r_a; a=1,\ldots,d$) do not commute with the
Schr\"odinger operator. Therefore, Barut argued that one may 
{\it ``\ldots fix $\cal M$, but change the transformation properties of $t$ 
and $\vec{r}$ in such a way that we obtain symmetry operations for the 
Schr\"odinger operator \ldots''} \cite{Baru73} and in this way, a contraction
$\mathfrak{conf}_{d+1}\to\mathfrak{sch}_d$ is claimed to be achieved. 
However, that procedure appears rather {\it ad hoc} and it might be useful 
to reconsider that derivation in somewhat more detail, treating $\cal M$ as
a dynamical variable from the outset and avoiding any ill-defined changes of
transformation properties.  

Starting again from (\ref{A:gl:1}), we define a new function 
$\chi(u,t,\vec{r})$ through
\BEQ
\vph_{\cal M}(t,\vec{r}) = \frac{1}{\sqrt{2\pi\,}} 
\int_{\mathbb{R}} \!\D u\, e^{-\II {\cal M}u}\, \chi(u,t,\vec{r}) 
\EEQ
which satisfies the equation of motion 
(if $\lim_{u\to\pm\infty}\chi(u,t,\vec{r})=0$)
\BEQ \label{A:gl:3}
\left( \frac{1}{c^2}\frac{\partial^2}{\partial t^2} +
\frac{\partial}{\partial \vec{r}}\cdot\frac{\partial}{\partial \vec{r}} 
+c^2 \frac{\partial^2}{\partial u^2}\right) \chi(u,t,\vec{r}) = 0
\EEQ
and the contact with the $(d+2)$-dimensional massless Klein-Gordon equation
and the associated conformal generators 
(\ref{3:gl:konfG}) is reached by defining $\Psi(\vec{\xi})=\chi(u,t,\vec{r})$
where $\xi_{-1}=u/c$, $\xi_0=ct$ and
$\xi_a=r_a; a=1,\ldots d$. Next, we define the wave function 
\BEQ
\psi(\zeta,t,\vec{r}) := \chi(u,t,\vec{r}) \;\; , \;\;
\zeta := u + \II c^2 t
\EEQ
which is the exact analogue of Barut's substitution (\ref{A:gl:subs}) 
in $\partial_t$ and have from (\ref{A:gl:3}) 
\BEQ
\left( 2\II \frac{\partial^2}{\partial\zeta\partial t} + 
\frac{\partial}{\partial \vec{r}}\cdot\frac{\partial}{\partial \vec{r}} 
\right) \psi(\zeta,t,\vec{r}) = 
\frac{1}{c^2}\frac{\partial^2}{\partial t^2}\psi(\zeta,t,\vec{r})
= {\rm O}\left( c^{-2}\right)
\EEQ
which reduces to the Schr\"odinger equation in the $c\to\infty$ limit. 
Next, we rewrite the generators of $\mathfrak{conf}_{d+2}$. For brevity,
we specialize to $d=1$. For the translations, we find
\BEQ
P_{-1} \Psi = -\II c M_0 \psi \;\; , \;\;
P_0    \Psi = c \left[ M_0\psi +{\rm O}\left(c^{-2}\right)\right] \;\; , \;\;
P_1    \Psi = - Y_{-1/2} \psi
\EEQ
For the rotations, we obtain
\BEQ
M_{01}\Psi = -c\left[ Y_{1/2}\psi+{\rm O}\left(c^{-2}\right)\right] \;\; , \;\;
M_{-11}\Psi= \II c\left[ Y_{1/2}\psi+{\rm O}\left(c^{-2}\right)\right] 
\;\; , \;\; 
M_{-10}\Psi=\II N\psi +{\rm O}\left(c^{-2}\right)
\EEQ
The dilatation becomes $D\Psi=(-2X_0+N)\psi$ and for the special conformal
transformations we find
\BEQ
K_{-1}\Psi = 2\II c \left[ X_{1}\psi+{\rm O}\left(c^{-2}\right)\right] 
\;\; , \;\; 
K_{0}\Psi = -2c \left[ X_{1}\psi+{\rm O}\left(c^{-2}\right)\right] 
\;\; , \;\; 
K_{1}\Psi =  -\left[ V_{+}\psi+{\rm O}\left(c^{-2}\right)\right] 
\EEQ 
Therefore, having treated the masses as dynamical variables from the
beginning, we rather have a projection of the complexified algebras
$(\mathfrak{conf}_{3})_{\C}\to\wit{\mathfrak{alt}}_1$
(and in general $(\mathfrak{conf}_{d+2})_{\mathbb{C}}
\to\wit{\mathfrak{alt}}_d$) when taking a non-relativistic
limit. We stress that the non-relativistic limit procedure actually 
throws out the time-translation operator $X_{-1}$. On the other hand, 
the operators $V_+$ and $N$ remain part of the algebra in the 
$c\to\infty$ limit. These results
are incompatible with Barut's claim that $\mathfrak{sch}_1$ would be obtained
in the non-relativistic limit, since time translations belong to the
Schr\"odinger group.

%%%%%%%%%%%%%%%%%%%%%%%%%%%%%%%%%%%%%%%%%%%%%%%%%%%%%%%%%%%%%%%%%%%%%%%%%%%%%%%%
\appsection{B}{ }
%%%%%%%%%%%%%%%%%%%%%%%%%%%%%%%%%%%%%%%%%%%%%%%%%%%%%%%%%%%%%%%%%%%%%%%%%%%%%%%%

We derive the causal representations eqs.~(\ref{5:gl:2}) and (\ref{5:gl:3P}). 
Because of rotation invariance and since the $\Psi_a$ are scalars, 
it is enough to consider the case $d=1$ explicitly. 
Beginning with the two-point function, we introduce for the phase
variables $\zeta_a$ center-of-mass coordinates $\eta=\zeta_1+\zeta_2$ and
relative coordinates $\zeta=\zeta_1-\zeta_2$. We then have
\BEA
\left\langle \phi_1 \phi_2^{*}\right\rangle &=&
\frac{1}{4\pi} \int_{\mathbb{R}^2}\!\D\zeta\D\eta\; 
e^{-\frac{\II}{2}({\cal M}_1-{\cal M}_2)\eta}
e^{-\frac{\II}{2}({\cal M}_1+{\cal M}_2)\zeta}
\left\langle \psi_1 \psi_2\right\rangle 
\nonumber \\
&=& \delta({\cal M}_1-{\cal M}_2) \int_{\mathbb{R}}\!\D\zeta\, 
e^{-\II {\cal M}_1\zeta} 
\left\langle \psi_1 \psi_2\right\rangle 
\nonumber \\
&=& \delta({\cal M}_1-{\cal M}_2) \delta_{x_1, x_2} \psi_0 t^{-x_1} 
{\cal M}_1^{x_1-1} 
\int_{\mathbb{R}} \!\D\zeta\, e^{-\II\zeta} \left( \zeta+\frac{\II{\cal M}_1}{2}
\frac{r^2}{t}\right)^{-x_1}
\nonumber \\
&=& \delta({\cal M}_1-{\cal M}_2) \delta_{x_1, x_2} \psi_0 t^{-x_1} 
{\cal M}_1^{x_1-1} \exp\left( -\frac{{\cal M}_1}{2}\frac{r^2}{t}\right) I
\EEA
where
\BEQ
I = \int_{\mathbb{R}+\II \frac{{\cal M}_1}{2}\frac{r^2}{t}} 
\!\D u\, u^{-x_1} e^{-\II u}
\EEQ
and $r=r_1-r_2$, $t=t_1-t_2$ and $\psi_0 = 4^{-x_1}\Psi_0$. The only 
singularity of the integrand is the cut along the negative real axis. Provided
the negative real axis is not crossed, the integration contour can be
arbitrarily shifted and therefore $I=I(x_1)$  depends {\em only} 
on the sign of $\frac{{\cal M}_1}{2}\frac{r^2}{t}$. 

Consider the case $\frac{{\cal M}_1}{2}\frac{r^2}{t}<0$, which implies
$t<0$ because of the physical convention ${\cal M}_1\geq 0$. Then the contour
of integration may be taken as $\mathbb{R}-\II \eps$ with $\eps>0$ and can be
closed by a semicircle in the lower half-plane. Using polar coordinates
$u=Re^{-\II\theta}$, the contribution $I_{\rm inf}$ 
of the lower semicircle of radius $R$ can be estimated in a standard fashion
\BEQ
\left| I_{\rm inf}\right| \leq 
R^{1-x_1} \int_{0}^{\pi}\!\D\theta\, e^{-R\sin\theta} \leq
2 R^{1-x_1} \int_{0}^{\pi/2} \!\D\theta\, e^{-(2R/\pi)\theta} \leq
\pi R^{-x_1}
\EEQ
and therefore vanishes as $R\to\infty$, provided $x_1>0$. 
It follows that $I=0$ for $t<0$ which proves (\ref{5:gl:2}). 

The three-point function is treated similarly. Introduce center-of-mass and
relative coordinates
\BEQ
\zeta = \zeta_1 - \zeta_3 \;\; , \;\; 
\zeta' = \zeta_2 - \zeta_3 \;\; , \;\; 
\eta = \zeta_1 + \zeta_2 + \zeta_3
\EEQ
Then
\BEA
\left\langle \phi_1 \phi_2 \phi_3^*\right\rangle &=& 
(2\pi)^{-3/2} \int_{\mathbb{R}^3} \!\D\zeta_1\D\zeta_2\D\zeta_3\; 
\exp\left( -\II{\cal M}_1\zeta_1-\II{\cal M}_2\zeta_2
+\II{\cal M}_3\zeta_3 \right) 
\left\langle \psi_1 \psi_2 \psi_3\right\rangle
\nonumber \\
&=& \frac{\delta({\cal M}_1+{\cal M}_2-{\cal M}_3)}{\sqrt{2\pi\,}} 
\int_{\mathbb{R}^2} \!\D\zeta\D\zeta'\; 
e^{-\II{\cal M}_1\zeta-\II{\cal M}_2\zeta'} 
\left\langle \psi_1 \psi_2 \psi_3\right\rangle
\EEA
Next, we set, using $r_{ab}=r_a-r_b$ and $t_{ab}=t_a-t_b$
\BEQ
u = \zeta + \II \frac{r_{13}^2}{2 t_{13}} \;\; , \;\;
u' = \zeta' + \II \frac{r_{23}^2}{2 t_{23}}
\EEQ
and
\BEQ \label{B:v}
v = \frac{r_{12}^2}{2 t_{12}}+ \frac{r_{23}^2}{2 t_{23}}
-\frac{r_{13}^2}{2 t_{13}} = 
\frac{1}{2} \frac{[r_{13}t_{23} - r_{23}t_{13}]^2}{t_{12}t_{23}t_{13}}
\EEQ
and find
\BEA
\left\langle \phi_1 \phi_2 \phi_3^*\right\rangle &=& 
\wit{\cal C}_{12,3} \delta({\cal M}_1+{\cal M}_2-{\cal M}_3)\, 
t_{12}^{-x_{12,3}/2} t_{23}^{-x_{23,1}/2} t_{13}^{-x_{13,2}/2}
\nonumber \\
& & \times \exp\left(-\frac{{\cal M}_1}{2}\frac{r_{13}^2}{t_{13}}  
-\frac{{\cal M}_2}{2}\frac{r_{23}^2}{t_{23}} \right)\: I
\EEA
where $\wit{\cal C}_{12,3} = \II C_{12,3}\,  2^{(x_1+x_2+x_3-1/2)}/\sqrt{\pi}$
and
\BEQ \label{B:eq:I3}
I = \int_{\mathbb{R}+\II\frac{r_{13}^2}{2t_{13}}} \!\D u\,
\int_{\mathbb{R}+\II\frac{r_{23}^2}{2t_{23}}} \!\D u'\; 
e^{-\II {\cal M}_1 u -\II {\cal M}_2 u'}\, 
(u-u'+iv)^{-x_{12,3}/2}\, {u'}^{-x_{12,3}/2}\, u^{-x_{13,2}/2}
\EEQ
Without restriction of the generality, we can take $t_{12}>0$. Otherwise, the
roles of $u$ and $u'$ in the following discussion will be exchanged. 
The integrand in $I$ has fixed cuts on the negative real axes of 
both $u$ and $u'$ and a movable cut which arises if $u-u'+\II v$ 
is real negative. From (\ref{B:v}) follows the important inequality 
\BEQ \label{B:eq:10}
v \geq \frac{r_{23}^2}{2 t_{23}} -\frac{r_{13}^2}{2 t_{13}}
\EEQ

%%==============================================================================
\begin{figure}
\epsfxsize=150mm
\centerline{\epsffile{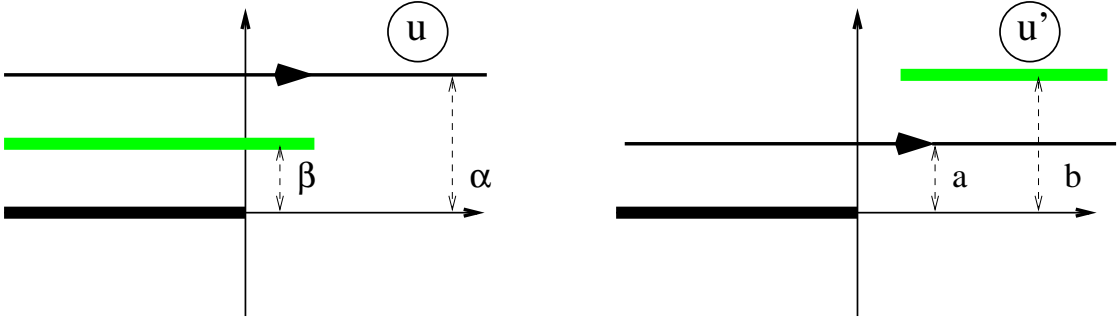}}
\caption{Integration contours in (\ref{B:eq:I3}) if $t_{23}>0$, indicated as 
oriented full lines. 
The thick black lines indicate the fixed cuts on the negative real axis 
and the grey lines indicate the moving
cuts which occur for $u-u'+\II v$ real negative. 
On the left the contour is shown
when one first integrates over $u$, with $\alpha =\frac{r_{13}^2}{2t_{13}}$
and $\beta=\frac{r_{23}^2}{2t_{23}}-v$. On the right one integrates first 
over $u'$, with $a=\frac{r_{23}^2}{2t_{23}}$ and 
$b=\frac{r_{13}^2}{2t_{13}}+v$.   
\label{Abb2}}
\end{figure}
%%==============================================================================

First, we consider the case $t_{23}>0$. Then $t_{13}>0$ as well and from
(\ref{B:v}) one also has $v\geq 0$. In figure~\ref{Abb2}, 
we show the integration contours in the complex plane,
when the integral over $u$ or $u'$, respectively, is performed first. Because
of the inequality (\ref{B:eq:10}), the contours never cross the moving cuts. 
Therefore, if one integrates first over $u$, the contour can be moved freely
in the upper half plane above the movable cut and consequently, $I$ must be
independent of $\frac{r_{13}^2}{2 t_{13}}$. On the other hand, if one 
integrates first over $u'$, the contour can be moved freely between the
singularities and $I$ is independent of $\frac{r_{23}^2}{2 t_{23}}$. Therefore
$I=I(v;{\cal M}_1,{\cal M}_2, x_1, x_2, x_3)$ and we have the integral
representation
\BEQ \label{B:eq:11}
I = \int_{\mathbb{R}+\II\eps}\!\D u\, \int_{\mathbb{R}+\II\eps'}\!\D u'\; 
e^{-\II{\cal M}_1 u-\II{\cal M}_2 u'}\, 
\left( u-u'+\II v\right)^{-x_{12,3}/2}\, {u'}^{-x_{23,1}/2}\, u^{-x_{13,2}/2}
\EEQ
with $\eps,\eps'\gtrsim 0$, at least if $v>0$. 

%%==============================================================================
\begin{figure}
\epsfxsize=80mm
\centerline{\epsffile{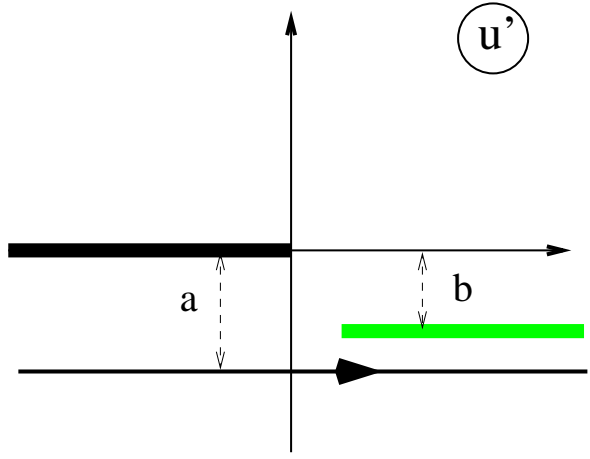}}
\caption{Integration contour in (\ref{B:eq:I3}) if $t_{23}<0$. 
The notation is the same as in figure~\ref{Abb2}. 
\label{Abb3}}
\end{figure}
%%==============================================================================

Second, we consider the case $t_{23}<0$. We shall show that $I=0$ provided
$x_2>0$.
It is convenient to carry out first the integration over $u'$. The contour
together with the cuts is shown in figure~\ref{Abb3}. The position of the
contour relative to the cuts follows from (\ref{B:eq:10}) and allows one
to close the contour in the lower half plane (the relative position of the cuts
depends on the values of the coordinates but does not matter).
We now show that the integral 
\BEQ
J = \int_{\R+\II\gamma} \!\D u'\, (\al+\II\beta -u')^{-x}
{u'}^{-y} e^{-\II u'}
\EEQ
vanishes for all real values of 
$\alpha,\beta$ and $\gamma$ such that $\gamma>\beta$ and if $x+y>0$. That
will imply $I=0$ under the stated conditions.  
In order to see this, consider the contour integral
$\oint_{\mathfrak{C}} \!\D u'\: \left( \al + \II \beta
            - u'\right)^{-x} {u'}^{-y} e^{-\II u'}$
The contour $\mathfrak{C}$ runs along
$\R+\II\gamma$ and is closed by a semi-circle
of radius $R$ in the lower half-plane. 
The condition $\gamma>\beta$ ensures that
the cuts lie on the outside of $\mathfrak{C}$ and therefore, the contour
integral vanishes. We introduce polar 
coordinates $u' = R e^{-\II \theta}$ and estimate the contribution
$J_{\rm inf}$ of the lower arc
\BEQ
\left| J_{\rm inf}\right| \leq R^{1-x-y} 
\int_{0}^{\pi}\!\D\theta\, e^{-R\sin\theta} B^{-x/2}
\EEQ
where $B$ is defined below. Because of the estimate
\BEQ
B := \left(1 -\frac{\al+\II \beta}{R}e^{\II\theta}\right) 
\left(1 -\frac{\al-\II \beta}{R}e^{-\II\theta}\right) 
\leq \left( 1 + \frac{|\al|}{R}\right)^2 + \left( 1+\frac{|\beta|}{R}\right)^2
\leq 3
\EEQ
which holds for $R$ sufficiently large, we have 
\BEQ
\left| J_{\rm inf}\right| \leq 3^{-x/2} R^{1-x-y} \int_{0}^{\pi}
\!\D\theta\, e^{-R\sin\theta} \leq 3^{-x/2}\pi R^{-x-y}
\EEQ
which tends to zero as $R\to\infty$ provided $x+y>0$, hence 
$J=0$. Finally, the exponents $x,y$ in $J$ are related to the 
scaling dimensions $x+y=(x_{12,3}+x_{23,1})/2=x_2>0$.  

In conclusion, the integral $I$ vanishes if $t_{12}>0$ and
$t_{23}<0$ and will therefore
contain a factor $\Theta(t_{23})$. By symmetry between $t_1$ and $t_2$,
the case $t_{12}<0$ will produce  a factor $\Theta(t_{13})$ provided
$x_1>0$. The scaling
function $\Phi_{12,3}(v)$ in eq.~(\ref{5:gl:3P}) can be identified with the
integral $I$ in (\ref{B:eq:11}) and the assertion is proven.

%%%%%%%%%%%%%%%%%%%%%%%%%%%%%%%%%%%%%%%%%%%%%%%%%%%%%%%%%%%%%%%%%%%%%%%%%%%%%%%%
\appsection{C}{On parabolic subalgebras}
%%%%%%%%%%%%%%%%%%%%%%%%%%%%%%%%%%%%%%%%%%%%%%%%%%%%%%%%%%%%%%%%%%%%%%%%%%%%%%%%

We  recall here the definition of parabolic subalgebras of a {\it complex}
simple Lie algebra $\g$, see \cite{Knap86}. 
The presentation is somewhat simpler than in 
the general case since for complex $\g$, the compact part and the
non-compact part of the Cartan decomposition
may be chosen to be the same up to multiplication by $\II$. 

Let $\g$ be a complex simple Lie algebra of rank $r$, 
$\h$ a complex Cartan subalgebra of $\g$ and $\Del$ the associated root
system. One chooses a basis of simple roots $\Pi=\{\al_1,\ldots,\al_r\}$
and denotes by $\Del_+\supset\Pi$ the
associated set of positive roots. If $\al\in\Del$ is a root, 
then $\g_{\al}\subset \g$
will be the corresponding root space. We shall also use the notation $\n=
\sum_{\al\in\Del_+}\g_{\al}$ for the subalgebra of $\g$ 
made up of all positive root spaces.

The  {\it minimal standard parabolic
subalgebra} $\s_0$ of $\g$ (associated with the given choice of $\h$ and 
$\Del_+$) is  defined as
\BEQ 
\s_0=\h\oplus\n=\h\oplus\sum_{\al\in\Del_+}\g_{\al}. 
\EEQ

A {\it standard parabolic subalgebra} of $\g$ is 
a subalgebra $\s\subset\g$ containing $\s_0$. 
More generally, a {\it parabolic subalgebra}
is defined to be a subalgebra of $\g$ containing a conjugate of $\s_0$.

The standard parabolic subalgebras $\s$ of $\g$
can be classified by means of a powerful result: they are in
one-to-one correspondence with the subsets $\Pi_{\s}$ of $\Pi$. Given $\Pi_{\s}
\subset \Pi$, here is how one constructs the associated parabolic subalgebra
$\s$. Let $\h_{\s}$ be the set of $H\in\h$ such that $\alpha(H)=0$ for all
$\alpha\in\Pi_{\s}$, or in other words the orthogonal in $\h$ of $\Pi_{\s}$; 
$\m_{\s}$ the centralizer of $\h_{\s}$
in $\g$, that is, the set of elements in $\g$ that commute with all elements
in $\h_{\s}$; finally $\n_{\s}\subset\n$ the sum of all 
root spaces associated with
positive roots $\al\in\Del_+$ that are not identically zero on $\h_{\s}$.
Then $\h_{\s}$, $\m_{\s}$ and $\n_{\s}$ are subalgebras of $\g$, with
$\h_{\s}\subset\h\subset\m_{\s}$, and
$\s$ is the direct sum 
\BEQ \label{C:gl:para}
\s=\m_{\s}\oplus\n_{\s}.
\EEQ

It is easily checked from the above definitions that eq.~(\ref{C:gl:para}) 
yields a Lie subalgebra of $\g$, and that  $\al\in\Pi$ is in $\Pi_{\s}$
if and only if $\g_{-\al}\subset\m_{\s}$ -- in which case $\g_{\al}$ is
also included in $\m_{\s}$. So, in particular, one also sees quite easily
that $\s$ is indeed a standard parabolic subalgebra. Since the standard
parabolic subalgebras are non-conjugate, it is straightforward to see how
this classification extends to a classification of all parabolic subalgebras.

The main motivation for the definition of parabolic subalgebras is that, if
$G$ is a  simple group with Lie algebra $\g$,
the pieces that appear in the Plancherel formula for $G$ all 
come from representations induced from certain standard 
parabolic subgroups of $G$ (integrating
standard parabolic subalgebras). 

The case $\g={\mathfrak{sl}}
(r+1,\C)$ is particularly illuminating. Take $\h$ to be
the space of diagonal matrices with vanishing trace. Let $\al_1=$diag
$(1,-1,0,\ldots,0),\ldots,\al_r=$diag$(0,\ldots,0,1,-1)$ form a basis $\Pi$
of the set of roots, with the usual identification of $\h$ with its dual. 
We choose a subset $\al_{i_1},\ldots,\al_{i_s}$ $(s\le r)$ of~$\Pi$. 
One says that $i$ {\it connects} with $j$ $(1\le i<j\le r+1)$ 
if $\al_i,\al_{i+1},\ldots,\al_{j-1}\in\Pi_{\s}$.
This defines the connected components of the set $\{1,\ldots,r+1\}$. Then 
diag$(x_1,\ldots,x_{r+1})\in\h_{\s}$ if and only if $x_1+\ldots+x_{r+1}=0$
and $x_i=x_j$ whenever $i$ and $j$
are connected. Therefore, a diagonal matrix  $H$ with vanishing trace
is  in $\h_{\s}$ if and only if its entries  situated in a single connected
component of $\{1,\ldots,r+1\}$ are all equal. So, $\m_{\s}$ is the set of 
block-diagonal matrices with vanishing trace that do not mix the different
components. 

For illustration, 
take the example $r=4$ and $\Pi_{\s}=\{\al_1,\al_3,\al_4\}$. An 
element $H\in\h_{\s}$ can be written as a diagonal matrix
\BEQ
H= \left(\begin{array}{ccccc} x_1\\ &x_1\\ &&x_2\\ &&&x_2\\ &&&&x_2
\end{array}\right) \;\; , \;\;
\mbox{\rm with $2x_1 + 3x_2=0$} 
\EEQ
In turn, elements $M\in\m_{\s}$ and $N\in\n_{\s}$ can be written as 
\BEQ
M=\left(\begin{array}{cc} M_1&0\\ 0&M_2\end{array} \right)\;\; , \;\;
N=\left(\begin{array}{cc} 0&N_1\\ 0&0\end{array} \right)
\EEQ
where $M_1,M_2$ and $N_1$ are, respectively, $2\times 2$, $3\times 3$ and
$2\times 3$ matrices and $\tr M_1 + \tr M_2=0$.

It is fairly easy to find the standard parabolic subalgebras in the case 
studied in section 3, namely $\g={\mathfrak{so}}(5,\C)$. 
Referring to the notations of that section,
the chosen basis of simple roots is
\BEQ 
\al_1=-e_2,\ \al_2=e_1+e_2. 
\EEQ
In table~\ref{tab1} we list all possible $\Pi_{\s},\h_{\s},\m_{\s},\n_{\s}$ 
and $\s$.

%%~~~~~~~~~~~~~~~~~~~~~~~~~~~~~~~~~~~~~~~~~~~~~~~~~~~~~~~~~~~~~~~~~~~~~~~~~~~~~~
\begin{table}
\caption{Construction of the standard parabolic subalgebras $\mathfrak{s}$ of 
the complex Lie algebra $\mathfrak{g}=(\mathfrak{conf}_3)_{\C}$.\label{tab1}}
\begin{center}
\begin{tabular}{|c|ccc|c|} \hline
$\Pi_{\s}$  & $\h_{\s}$  & $\m_{\s}$   & $\n_{\s}$ & $\s$ \\ \hline
$\emptyset$ & $\h$       & $\h$        & $\n$      & $\wit{\mathfrak{age}}_1$\\
$\{\al_1\}$ & $\C N$     & $\h\oplus\C Y_{-\half}\oplus\C V_+$ & 
$\C X_1\oplus\C M_0\oplus\C Y_{\half}$ & $\wit{\mathfrak{alt}}_1$ \\
$\{\al_2\}$ & $\C (N-D)$ & $\h\oplus\C X_{-1}\oplus \C X_1$ & 
$\C Y_{-\half}\oplus \C Y_{\half} \oplus \C M_0$ & $\wit{\mathfrak{sch}}_1$ \\
$\{\al_1,\al_2\}$ & $\{0\}$ & $\g$ & $\{0\}$ & $(\mathfrak{conf}_3)_{\C}$\\ 
\hline
\end{tabular} \end{center}
\end{table}
%%~~~~~~~~~~~~~~~~~~~~~~~~~~~~~~~~~~~~~~~~~~~~~~~~~~~~~~~~~~~~~~~~~~~~~~~~~~~~~~

Let us explain how to find the entries in table~\ref{tab1} 
in the case where $\Pi_{\s}=\{\al_1\}=\{-e_2\}$. Then
\BEQ
\h_{\s}=\{\lambda N+\mu D \ |\ \al_1(\lambda N+\mu D)=0\}=\{\lambda N\}
=\C N
\EEQ
where we recall from section~3 that $e_i(N)=\delta_{i,1}$ and 
$e_i(D)=\delta_{i,2}$. So
\BEA
\m_{\s} &=& \{X\in\g\ |\ [X,\h_{\s}]=0 \} = \h\oplus \g_{-e_2}\oplus
\g_{e_2} 
\nonumber \\
        &=& \h\oplus \C Y_{-\half}\oplus \C V_+.
\EEA
To find $\n_{\s}$, we must determine all positive roots which 
vanish on $\h_{\s}$. In this case, the only such positive root is $-e_2$. 
Consequently, with the set $\Delta_+$ of
positive roots given by (\ref{3:gl:posWur})
\BEQ
\n_{\s}=\g_{e_1+e_2}\oplus\g_{e_1}\oplus\g_{e_1-e_2} 
       =\C X_1\oplus\C Y_{\half}\oplus\C M_0.
\EEQ

%%%%%%%%%%%%%%%%%%%%%%%%%%%%%%%%%%%%%%%%%%%%%%%%%%%%%%%%%%%%%%%%%%%%%%%%%%%%%%%%

\end{document}